\documentclass[pre,aps,twocolumn,amsmath,amssymb,superscriptaddress]{revtex4}

\pdfoutput=1
\usepackage{hyperref}
\usepackage{graphicx}
\usepackage{bm}
\usepackage{color,soul}
\usepackage{amsmath}
\usepackage{amssymb}

\def\beq{\begin{equation}}
\def\eeq{\end{equation}}
\def\bea{\begin{eqnarray}}
\def\eea{\end{eqnarray}}

\begin{document}

\title{Stochastic level-set method for shape optimisation}

\author{Lester O. Hedges}
\affiliation{Department of Physics, University of Bath, Bath BA2 7AY, United Kingdom}
\affiliation{School of Engineering, Cardiff University, Cardiff CF24 3AA, United Kingdom}
\author{H. Alicia Kim}
\affiliation{Structural Engineering Department, University of California, San Diego, La Jolla, CA 92093 USA}
\affiliation{School of Engineering, Cardiff University, Cardiff CF24 3AA, United Kingdom}
\author{Robert L. Jack}
\affiliation{Department of Physics, University of Bath, Bath BA2 7AY, United Kingdom}

\begin{abstract}
We present a new method for stochastic shape optimisation of engineering structures.  The method generalises an existing deterministic scheme, in which the structure is represented and evolved by a level-set method coupled with mathematical programming.  The stochastic element of the algorithm is built on the methods of statistical mechanics and is designed so that the system explores a Boltzmann-Gibbs distribution of structures.  In non-convex optimisation problems, the deterministic algorithm can get trapped in local optima: the stochastic generalisation enables sampling of multiple local optima, which aids the search for the globally-optimal structure.  The method is demonstrated for several simple geometrical problems, and a proof-of-principle calculation is shown for a simple engineering structure.
\end{abstract}
\maketitle

\newcommand{\rlj}[1]{{\color{blue} #1}}
\renewcommand{\rlj}[1]{{#1}}

\newcommand{\xx}{\bm{x}}
\newcommand{\RR}{\mathbb{R}}
\newcommand{\XX}{\bm{X}}
\newcommand{\vn}{v^{\rm n}}
\newcommand{\Vn}{V^{\rm n}}
\newcommand{\eps}{\epsilon}

\section{Introduction}

Structural optimisation aims to provide performance improvements and/or weight savings by formulating an engineering structural design problem as constrained optimisation. The class of structural optimisation that is of interest in this manuscript is shape optimisation using the level-set method, which systematically modifies the structural boundary, i.e. design shape, to maximise or minimise the given performance metric while satisfying one or more constraints. A related class of structural optimisation is topology optimisation in which shapes and the number of boundaries are optimised. Topology optimisation is considered the most generic form of structural optimisation since the optimal solution is the most independent of the initial solution and can offer substantial performance improvements via unintuitive and creative designs ~\cite{Sigmund2001,Deaton2014}. To this extent, shape optimisation is an important class of structural optimisation, critically enabling topology optimisation via the level-set method. It is noted that the level-set method naturally splits or merges boundaries thus topology optimisation is inherently enabled, although for the careful investigations presented in this article, we focus our attention to shape optimisation of a single external boundary of the structure.

It is well-known that many of the relevant applications involve optimisation problems that are \emph{non-convex} -- they support multiple locally-optimal designs, which correspond to local minima of the objective function.  These local optima might be associated with different structural topologies, conflicting design requirements and/or with numerical aspects of the (discretised) computational problem. Examples of non-convex design spaces in the engineering literature are stress constrained optimization~\cite{brampton} and coupled multiphysics optimization~\cite{Dunning2015-aero}.  Such systems cause problems for conventional (deterministic) optimisation schemes, which tend to converge to \emph{local} optima, but miss the globally-optimal structure.  The purpose of this paper is to exploit an analogy between such engineering problems and statistical mechanical systems with non-convex potential energy surfaces.  This motivates us to introduce a stochastic algorithm for exploring the space of possible structures, in order to sample multiple local optima and -- eventually -- converge to the global optimum.

In the engineering context, non-convexity and multiple optima present interesting dilemmas.  If optimisation algorithms yield locally optimal structures which are much worse than the global optimum, they would not be considered useful design methods.  However, generating multiple locally-optimal solutions can also provide multiple design ideas to engineers, offering a range of possible solutions with similar values of the objective function.  In this case, an engineer may wish to consider several of these solutions, based on practical design requirements such as ease of manufacturing. This particularly true when there are several designs of similar objective function and constraint values, as in~\cite{brampton}.

A common approach to topology optimisation is to employ gradient-based nonlinear programming, which can be applied to typical engineering design problems with $10^4 - 10^6$ design variables~\cite{bendsoe,Dunning2015}.  In this case, iteration of the optimiser leads to one local optimum solution. Alternative (stochastic) approaches such as evolutionary algorithms, particle swarm optimisation and simulated annealing are capable of searching for multiple potential solutions and they have been applied to topology optimisation~\cite{Wang2005,Wu2010,Luh2011}. However the success of such methods has been limited, partly because they do not typically take advantage of information about the gradient of the objective function. Interested readers are referred to a critical review~\cite{Sigmund2011} which presents an example: Topology optimisation was applied to a small problem with just 144 variables, which was solved using the non-gradient method of differential evolution. This required 15,730 function evaluations. In contrast, a gradient-based topology optimisation method -- Solid Isotropic Material with Penalisation (SIMP) -- converged to a slightly superior solution after just 60 function evaluations. This motivates the research question of how to explore non-convex engineering design spaces effectively and efficiently. 

One approach which can be successful in non-convex optimisation problems is to start with a deterministic optimisation procedure and to add a  stochastic component, so that the objective function can both increase and decrease as the algorithm runs.  In this article, we introduce a stochastic optimisation procedure which aims to generate designs according to a prescribed distribution, analogous to the Gibbs-Boltzmann distribution of statistical mechanics.  The method is built on the deterministic gradient-based topology optimisation method of Dunning and Kim~\cite{Dunning2015,Siva2016} and that method is recovered in the zero-noise limit of the stochastic process -- this means that the stochastic method should perform at least as effectively as the deterministic one.  Moreoever, since the model is based on the Boltzmann distribution, we expect that it can be combined with parallel-tempering methods~\cite{frenkelsmit,Sugita1999,Rosta2009}, which offer a systematic approach for exploring a range of near-optimal structures.

Our new method is based the level-set topology optimization (LSTO) method of Refs.~~\cite{Dunning2015,Siva2016}.
Unlike traditional topology optimisation (e.g. SIMP) in which the design variables (for example $n$) indicate whether each element of the design domain should exist ($n=1$) or be absent ($n=0$), the level-set method employs an implicit functional representation $\phi$ that directly represents the domain boundaries. A key advantage of this method is that topological changes in the shape of the object (for example the removal of holes) are associated with singularities in the behaviour of the boundary of the object, but do not involve any singularities in $\phi$.  See~\cite{OsherBook} for a description of the level-set method and its key advantages.  The effect of the level-set representation is that topology optimisation can be reformulated as an extended shape optimisation, where boundary shapes are optimised and the number of boundaries can change. In this paper, we will focus on the shape optimisation aspect of the algorithm, where the level set method moves the boundaries using an advection equation. Within this scheme, the LSTO method corresponds to steepest descent of the objective function, which ensures robust convergence of the method to a locally-optimal design.  We will show how this method can be extended to a stochastic method that can explore non-convex design landscapes. 

This article presents several new results.  In Sec.~\ref{sec:level-set}, we review the method of~\cite{Dunning2015}, and we introduce some simplifications to that method, which clarify the relationship between that method and steepest descent optimisation of the objective function.  In Sec.~\ref{sec:stochastic}, we introduce a stochastic process that explores a range of structures, parameterised by a noise strength $T$ that is analogous to the temperature in statistical physics.  The process is defined by a stochastic differential equation, analogous to Langevin equations in physics.  Sec.~\ref{sec:results} includes several examples of the application of the stochastic method, including matching of a shape to a fixed design, and compliance minimisation of a simple two-dimensional engineering structure, subject to a constraint on its area.  We discuss the nature of the shapes/structures explored by the stochastic method, and discuss the potential use of the method for practical optimisation of engineering structures.   Sec.~\ref{sec:conclusions} summarises our conclusions.

\section{The level set topology optimisation method}
\label{sec:level-set}

\begin{figure*}[tb]
\centering
\includegraphics[width=14cm]{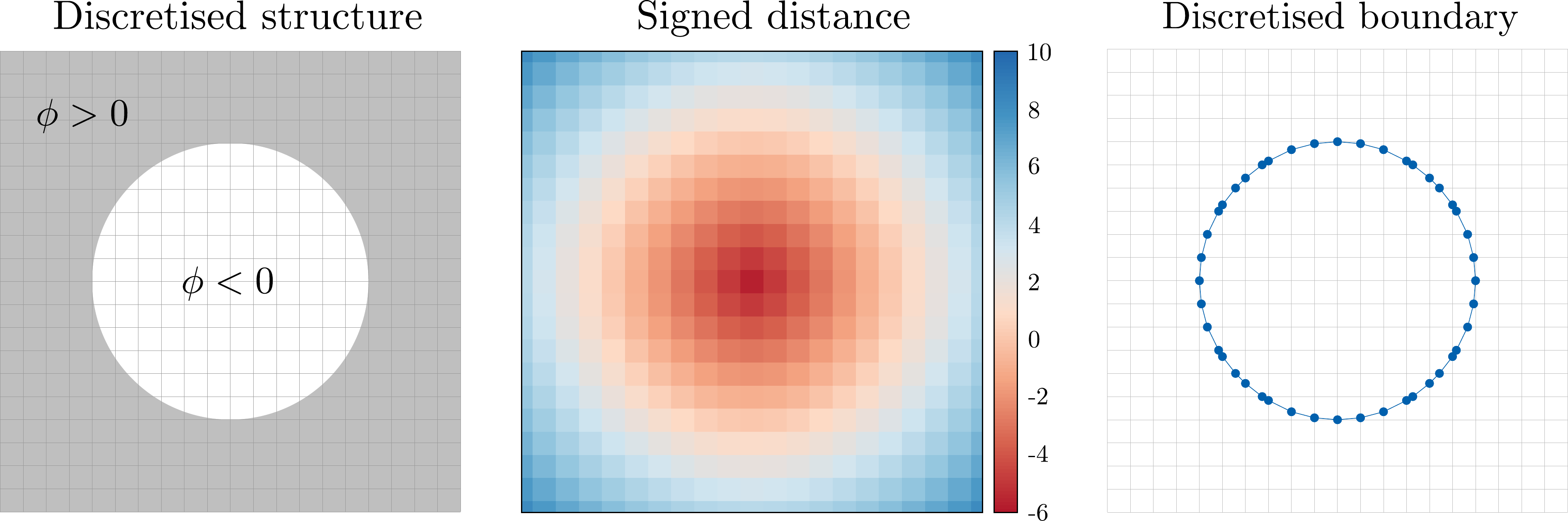}
\caption[Illustration of the level-set construction]{\label{fig:level-set} The level-set domain: (Left) A  two-dimensional square domain $\Omega_{\rm d}$ is discretised using a uniform square grid. A function $\phi$ defined on the nodes of the grid indicates whether each node is inside ($\phi < 0$) or outside ($\phi > 0$) a structure $\Omega$.  \rlj{In this case $\Omega$ is a square that contains a circular hole.} (Middle) The function $\phi$ is given by the signed distance to the nearest point on the boundary of $\Omega$. With this choice $|\nabla \phi| = 1$.  (Right) A discretised representation of the boundary of $\Omega$ is obtained by defining boundary points, which are located either on nodes (if $\phi_i=0$) or on edges between nodes (if this edge connects two nodes with opposite signs for $\phi$). The resulting set of boundary points provide a piecewise linear approximation for of the boundary of $\Omega$.}
\end{figure*}

The optimisation problem considered here is the determinisation of a structural domain $\Omega$ that minimises an objective function $F(\Omega)$, subject to some constraints.  
This section introduces the computational method used here, which is closely based on the LSTO method of Dunning and Kim~\cite{Dunning2015}, see also~\cite{Siva2016}.  
The implementation described here is a slightly simplified version of the original LSTO method: our C++ implementation is available at~\cite{slsm}.  We give a brief and informal description of this (deterministic) optimisation algorithm, to set the scene for the stochastic method that we will describe in Sec.~\ref{sec:stochastic}.  Rigorous discussions of the properties of the level-set method and of the mathematical results underlying this algorithm can be found elsewhere~\cite{OsherBook}.

To define our optimisation problem, we require an objective function $F$, a constraint function $G$, and a design domain $\Omega_{\rm d}\subset \RR^d$, where $d$ is the spatial dimensionality (In this work we take $d=2$ although generalisation to higher dimension is possible.)  The aim of our optimisation is find a structure $\Omega \subset \Omega_{\rm d}$ such that $F(\Omega)$ is minimised, subject to the constraint  
\begin{equation} 
  G(\Omega)\leq G^* . \label{equ:GG} 
\end{equation}  
Extension to multiple constraints or equality constraints of the form $G(\Omega)=G^*$ is straightforward~\cite{Dunning2015,Siva2016} but we consider a single inequality constraint here, for simplicity.  The LSTO method prescribes a time-evolution for $\Omega$ that converges to a (local) optimum of $F$, which satisfies the constraint.

\subsection{Evolution in continuous space and time}
\label{subsec:cts-det}

The structure $\Omega$ is defined in terms of a real-valued function $\phi : \Omega_{\rm d}\to \RR$. That is, define $\Omega = \{\bm{x} \in \Omega_{\rm d}: \phi(\bm{x})\geq 0\}$ as the part of the domain for which $\phi\geq0$.  Throughout this work, bold symbols such as $\bm{x}$ indicate vectors in $\RR^d$.  The boundary of $\Omega$ is denoted by $\Gamma$ and is defined as the zero level-set of $\phi$, that is $\Gamma = \{ \bm{x} \in \Omega_{\rm d}: \phi(\bm{x})= 0\}$.   

The boundary $\Gamma$ is made up from one or more closed curves, so we index the points in $\Gamma$ by an internal co-ordinate $u>0$, such that $\XX(u)\in \Gamma$ is a point on the boundary of $\Omega$.  Also let $\bm{n}(u)$ be the (\rlj{inward}) normal vector to the boundary at the point $\XX(u)$, and define a function $\ell'$ such that $\int_{u_1}^{u_2} \mathrm{d}\ell(u)=\int_{u_1}^{u_2}\ell'(u)\mathrm{d}u$ is the length of the boundary between the two points $\XX(u_1)$ and $\XX(u_2)$.

Several level-set topology optimisation approaches~\cite{bendsoe,Deaton2014} use a steepest descent strategy for the minimisation of $F(\Omega)$.  To achieve this, let the function $\phi$ evolve as a function of the time $t$ according to an advection equation
\begin{equation}
\frac{\partial}{\partial t} \phi(\xx,t) = -\bm{v}(\xx,t)\cdot\nabla\phi(\xx,t)
\label{equ:advect}
\end{equation}
where $\bm{v}$ is a local velocity, to be specified below.  Assume that for $t=0$ then $\phi$ solves the eikonal equation $|\nabla\phi|=1$.  Given that $\phi=0$ on the boundary $\Gamma$, this is equivalent to taking $\phi(\xx)$ to be the (signed) distance of point $\xx$ from this boundary.  The vector $\nabla\phi$ is normal to the level sets of $\phi$, and one sees from (\ref{equ:advect}) that the time-evolution of $\phi$ depends only on the normal velocity \rlj{$\vn=\bm{v}\cdot\nabla\phi/|\nabla\phi|$}.  In what follows we take $\nabla\vn\cdot\nabla\phi=0$ which ensures that if the eikonal equation is true at $t=0$, then $|\nabla\phi|=1$ for all times $t>0$.  \rlj{Note that for points on the boundary $\Gamma$, the inward normal is $\bm{n}(u)=\nabla\phi(\xx(u))/|\nabla\phi(\xx(u))|$.}

Hence, to specify the time evolution of $\phi$, it is sufficient to specify $\vn$ for all points on the boundary $\Gamma$ of the structure $\Omega$, since the condition $\nabla\vn\cdot\nabla\phi=0$ then specifies $\vn$ at all other points.   In order that (\ref{equ:advect}) corresponds to steepest descent for the objective function $F$, we introduce a sensitivity $s^F$, so that $s^F(u)$ is the sensitivity for $F$ at the point $\XX(u)$.  To define $s^F$, consider a deformed structure $\Omega_\eps$ whose boundary $\Gamma_\eps$ consists of points $\XX_\eps(u)=\XX(u)+\eps z(u)\bm{n}(u)$ where the function $z$ sets the size of the displacement of each boundary point.  Informally, the sensitivity $s^F(u)$ is the rate of change of $F$ associated with moving the boundary point $\XX(u)$ in the direction $\bm{n}(u)$.  More precisely, $s^F$ is the unique function that obeys
\begin{equation}
F(\Omega_\eps) = F(\Omega) + \eps \int_\Gamma s^F(u) z(u) \mathrm{d}\ell(u) + O(\eps^2) .
\label{equ:sensF}
\end{equation}
We assume in the following that the objective function $F$ and the boundary $\Gamma$ are sufficiently smooth that this sensitivity function exists.  For a rigorous discussion of these issues, see~\cite{Allaire2004}.

To perform an unconstrained minimisation of $F(\Omega)$, one should prepare an initial condition in which $\phi$ is the signed distance from the boundary of some initial design $\Omega_0$.  Then one should solve (\ref{equ:advect}), with the normal velocity at boundary point $\XX(u)$ being 
\begin{equation}
\vn(\XX(u))=- s^F(u) .
\label{equ:vnsf}
\end{equation}
The normal velocity $\vn$ at any point $\xx$ which is not on the boundary $\Gamma$ should be set equal to the normal velocity at the nearest point on $\Gamma$, which ensures that $\nabla \vn\cdot \nabla\phi=0$, as noted above.  This time-evolution for $\phi$ encodes a time-evolution for the structure $\Omega$: given that $\phi$ evolves in this way, it is easy to verify that $\partial_t F(\Omega) = -\int s^F(u)^2 \mathrm{d}\ell(u) \leq 0$, so the objective function decreases with time.

We now consider minimisation of $F(\Omega)$ subject to a constraint,  that $G(\Omega)\leq G^*$.  To achive this, a Lagrange multiplier $\mu^G$ is introduced~\cite{Siva2016}, such that
\begin{equation}
\vn(\XX(u))=  -s^F(u) + \mu^G s^G(u) 
\label{equ:vn-lambda-cts}
\end{equation}
where $s^G$ is the sensitivity for the constraint function $G$, and $\mu^G$ is chosen (independent of $u$) such that the system does not violate the constraint on $G$.  The generalisation of this construction to systems with multiple constraints or to equality constraints of the form $H(\Omega)=0$ is straightforward but we restrict here to just one constraint, for compactness of notation.

\subsection{Discrete space and time, and boundary discretisation}
\label{sec:discrete-det}

For a computational implementation, these equations must be discretised in both space and time.  For the spatial discretisation, the domain $\Omega_{\rm d}$ is partitioned into a square grid of side $a_0=1$: see Fig.~\ref{fig:level-set}.  The vertices of the grid are called nodes.  Each node $i$ has a position $\xx_i$ and an associated value of the level-set function $\phi_i=\phi(\xx_i)$.  The temporal discretisation is a simple first-order Euler scheme, so (\ref{equ:advect}) becomes
\begin{equation}
\phi_i(t+\Delta t) = \phi_i(t) - \vn_i(t)\Delta t\cdot |\nabla\phi(t)|_i
\label{equ:advect-discrete}
\end{equation}
where $\vn_i$ is the normal velocity at node $i$ and $|\nabla\phi(t)|_i$ is the modulus of $\nabla\phi$, evaluated at $\bm{x}_i$, which is equal to unity if $\phi$ solves the eikonal equation exactly.  In practice, $\nabla\phi$ is estimated for each node using the Hamilton-Jacobi weighted essentially non-oscillatory method (HJ-WENO)  described in~\cite{OsherBook}.

\begin{figure}[tb]
\centering
\includegraphics[width=7cm]{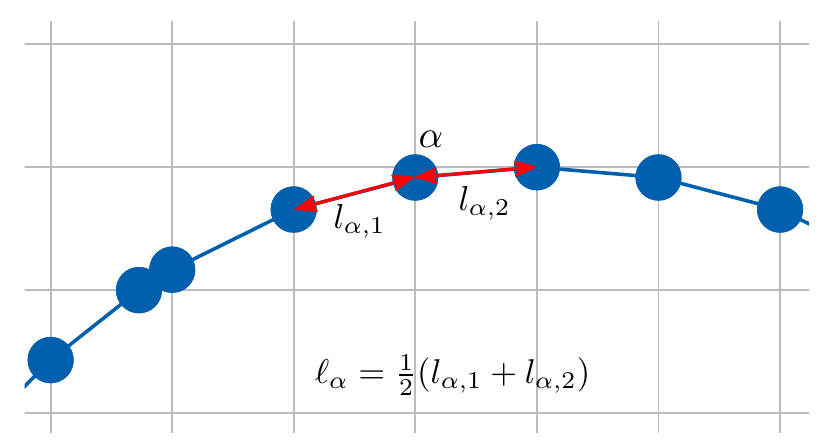}
\caption[Illustration of boundary discretisation]{\label{fig:boundary-integral} Enlarged view of the discretised zero contour of the level set function. Each boundary point $\alpha$ is associated with two segments of the piecewise linear boundary. For each boundary point, we define an associated boundary length $\ell_\alpha = \left(l_{\alpha,1} + l_{2,\alpha,2}\right)/2$.}
\end{figure}

In order to determine the normal velocities $\{\vn_i\}$, the LSTO method employs a second level of discretisation: see Fig.~\ref{fig:boundary-integral}.   From the (discretised) level set function $\phi$, a (discrete) set of boundary points is inferred, as follows: If $\phi_i=0$ then node $i$ is a boundary point.  Also, if two adjacent nodes have $\phi$-values with opposite signs then there is a boundary point between them, which is taken to lie on the edge between the nodes, with a position determined by linear interpolation.  Let the number of boundary points be $n$ and let the position of boundary point $\alpha$ be $\XX_\alpha$ (with $1\leq\alpha\leq n$).  The boundary points form a set of closed curves, which provide a discrete representation of the boundary $\Gamma$.  

The reason for this boundary discretisation is that the LSTO method uses estimates of the sensitivities $s^F,s^G$ on the boundary points (this is the natural choice since the sensitivity is intrinsically related to the boundary $\Gamma$).  Given these sensitivities, one infers a velocity $V_\alpha^{\rm n}$ for each boundary point $\alpha$: these velocities are determined by a linearised optimisation sub-problem that is solved at each time step.  From the boundary point velocities $V_\alpha^{\rm n}$, the normal velocities $\vn_i$ for each node are calculated by a fast-marching method. Hence the level set variables can be propagated forward in time, according to (\ref{equ:advect-discrete}).  We now describe these steps in more detail.

\subsection{Determination of boundary point velocities from an optimisation sub-problem}
\label{sec:boundary-vel-opt}

To determine the boundary point velocities $\Vn_\alpha$, we suppose that estimates of the sensitivities $s^F$ and $s^G$ are available for each boundary point.  For boundary point $\XX_\alpha$ let these estimates be $s^F_\alpha,s^G_\alpha$.  (Estimation of these sensivities depends on the problem of interest and will be discussed in Sec.~\ref{sec:results}.)  Also define $l_{\alpha,1}$ and $l_{\alpha,2}$ as the distances from point $\alpha$ to its neighbouring boundary points; and let $\ell_\alpha=(l_{\alpha,1}+l_{\alpha,2})/2$ be the length of boundary associated with point $\alpha$, as in Fig.~\ref{fig:boundary-integral}.  (It follows that $\sum_{\alpha=1}^n \ell_\alpha$ is the total boundary length.)  Now suppose that each boundary point $\XX_\alpha$ moves a distance $\Vn_\alpha\Delta t$ in the normal direction: discretising (\ref{equ:sensF}) along the boundary, the change in the objective and constraint functions can be estimated as
\begin{equation}
    \begin{split}
        \Delta F &= \sum_{\alpha=1}^n \Vn_\alpha s^F_\alpha \ell_\alpha \Delta t,\\
        \Delta G &= \sum_{\alpha=1}^n \Vn_\alpha s^G_\alpha \ell_\alpha \Delta t.
    \end{split}
    \label{equ:dFG}
\end{equation}
Direct optimisation of the $\Vn_\alpha$ can then be used to optimise the change in $F$, given any constraints (see for example~\cite{Dunning2015}).

However, optimising over all the parameters $\Vn_\alpha$ is not convenient numerically: there is a large number of such parameters, and discretisation errors can result in rough boundaries $\Gamma$~\cite{Dunning2015}.  Instead, the LSTO method uses the (constrained) steepest-descent defined by (\ref{equ:vn-lambda-cts}), with a variable time step that is optimised according to the values of the sensitivities.  Spatial discretisation of (\ref{equ:vn-lambda-cts}) yields
\begin{equation}
\Vn_\alpha \Delta t=  \lambda^F s^F_\alpha  + \lambda^G s^G_\alpha,
\label{equ:vn-lambda-disc}
\end{equation}
where $\lambda^F,\lambda^G$ are parameters to be determined (with $-\lambda^F$ corresponding to the time step $\Delta t$ in (\ref{equ:vn-lambda-cts}) and $\lambda^G$ corresponding to $\mu^G \Delta t$). 

At each iteration, the LSTO method optimises the parameters $\lambda^F,\lambda^G$, in order to make $\Delta F$ as negative as possible.  However, this procedure is subject to several constraints, which include both the optimisation constraint (\ref{equ:GG}), and several additional considerations, which are determined by the physical characteristics of the topology optmisation problem.  Note that the handling of these constraints in the algorithm presented here differs from the method of~\cite{Dunning2015}.  The method presented here has been chosen to combine simplicity and accuracy.

First, note that $\Delta F,\Delta G$ are \emph{estimates} for the changes in $F,G$, which are accurate only if the boundary point displacements are not too large.  To avoid very large values of the $\lambda$ parameters, a 
constraint 
is applied to each boundary point displacement:
\begin{equation}
|\Vn_\alpha \Delta t| \leq d_{\rm CFL}
\label{equ:cfl}
\end{equation}
Note that consistency between (\ref{equ:vn-lambda-disc}) and (\ref{equ:vn-lambda-cts}) shows that $-\lambda^F$ is equal to the timestep $\Delta t$: the primary role of this constraint is to impose a sufficiently small time step for the Euler discretisation (\ref{equ:advect-discrete}).  To determine an appropriate value for $d_{\rm CFL}$, we use the Courant-Friedrichs-Levy (CFL) condition associated with (\ref{equ:advect}), which requires that $d_{\rm CFL}$ be smaller than the grid spacing.  Large $d_{\rm CFL}$ leads to larger time steps and hence faster optimisation, but if the resulting convergence histories are interpreted as as approximate solutions to (\ref{equ:advect}) then these solutions are less accurate when $d_{\rm CFL}$ is large.  The specific value of $d_{\rm CFL}$ depends on the problem of interest.

To implement the CFL constraint, 
the optimisation domain for $\lambda^F$ is $|\lambda^F|\leq\lambda^F_{\rm max}$ with $\lambda^F_{\rm max}=-d^{\rm CFL}/(\max_\alpha|s^F_\alpha|)$.  A similar restriction is applied to $\lambda^G$.  Since these constraints are applied separately to $\lambda^{F}$ and $\lambda^G$, the resulting solution may still violate the constraint (\ref{equ:cfl}): if this happens then the resulting $\lambda$ parameters are rescaled by a factor $ d_{\rm CFL}/(\max_\alpha|\Vn_{\alpha} \Delta t|)$, and the $\Vn_\alpha$ are recalculated using (\ref{equ:vn-lambda-disc}), so that (\ref{equ:cfl}) is then satisfied.  

Second, if the design at time $t$ is $\Omega_t$, the inequality constraint (\ref{equ:GG}) for the optimisation requires $G(\Omega_t) + \Delta G \leq G^*$.  If $\Omega_t$ does not satisfy the constraint ($G(\Omega_t)>G^*$) then the optimiser may not be able to find any solution for which $\Delta G$ is sufficiently negative to solve the constraint.  For this reason, the optimisation is performed subject to a modified constraint $\Delta G \leq -G_0$ where $G_0=G(\Omega_t)-G^*$ if a solution is possible, otherwise a smaller value for $G_0$ is chosen.  (In practice, the smallest (most negative) possible value for $\Delta G$ is calculated by considering the cases $\lambda^{F,G}=\pm \lambda^{F,G}_{\rm max}$: if the most smallest possible value $\Delta G^{\rm min}$ is greater than $G^*-G(\Omega_t)$ then it is likely that the constraint cannot be satisfied so we take $G_0=-c\Delta G^{\rm min}$, where the parameter $c$ can be adusted according to the problem of interest. Typically we take $c\approx 0.5$.  Note that $\Delta G^{\rm min}>G^*-G(\Omega_t)$ usually happens only during the initial stages of optimisation, so the way that this case is handled does not typically affect the convergence of the method to its final solution.)

Third, note that boundary points should not move outside the design domain $\Omega_{\rm d}$.  To avoid this problem, let $d_\alpha$ be the signed distance of  boundary point $\alpha$ from the boundary of $\Omega_{\rm d}$.  (The sign of $d_\alpha$ is positive if the normal vector $\bm{n}_\alpha$ points towards the domain boundary, negative otherwise.)  We estimate that point $\alpha$ has moved outside the domain if $\Vn_\alpha \Delta t>d^\alpha>0$ or $\Vn_\alpha \Delta t<d^\alpha<0$ (this is an approximation since the normal velocity is not perpendicular to the interface, but it is sufficient for our purposes).  If this happens we replace (\ref{equ:vn-lambda-disc}) by $\Vn_\alpha\Delta t = d_\alpha$.

With these ingredients in place, we finally define the full optimisation problem that is used to determine $\lambda^F,\lambda^G$.  Combining (\ref{equ:dFG},\ref{equ:vn-lambda-disc}), and accounting for the possibility that boundary points might move outside of the domain $\Omega_{\rm d}$, define
\begin{equation}
\Delta \hat{F}(\lambda^F,\lambda^G) = \sum_{\alpha=1}^n s^F_\alpha z_\alpha \ell_\alpha , \quad \Delta \hat{G}(\lambda^F,\lambda^G) = \sum_{\alpha=1}^n s^G_\alpha z_\alpha \ell_\alpha
\label{equ:FGhat}
\end{equation}
where $z_\alpha=\lambda^F s^F_\alpha  + \lambda^G s^G_\alpha$ if the boundary point remains inside the domain, and $z_\alpha=d_\alpha$ otherwise.  We then choose $\lambda^{F,G}$ to minimise $\Delta \hat{F}(\lambda^F,\lambda^G)$ subject to $ \Delta \hat{G}(\lambda^F,\lambda^G)\leq G_0$, on the domain $|\lambda^F|<\lambda^F_{\rm max},|\lambda^G|<\lambda^G_{\rm max}$.  Given the $\lambda^{F,G}$ solving the optimisation problem, the time step is $\Delta t = -\lambda^F$ and the boundary point velocities are $\Vn_\alpha = z_\alpha/\Delta t$.  

In practice, this two-parameter optimisation sub-problem is solved using the SLSQP method from the NLOPT package~\cite{nlopt}.  The simple form of (\ref{equ:FGhat}) means that derivatives of $\Delta \hat{F},\Delta \hat{G}$ can be obtained analytically without the need for the finite differences used in~\cite{Dunning2015}.  If some sensitivities are very large, it is convenient to precondition the optimiser by defining rescaled parameters $\tilde\lambda^F=\lambda^F/A$ and corresponding rescaled sensitivities $\tilde s^F=As^F$ for some parameter $A$, to ensure efficient solution of this sub-problem.

\subsection{Level set update}
\label{sec:levelset-update}

Having calculated the boundary point velocities $\Vn_\alpha$, the velocities $\vn_i$ on nodes adjacent to the boundary points are calculated by inverse-squared distance weighting, as in~\cite{Dunning2011}.  That is, if the edges associated with node $i$ of the grid include $m$ boundary points, then these points are indexed by $\beta=1\dots m$. Let the velocity of point $\beta$ be $\Vn_\beta$ and its distance from node $i$ be $r_\beta$.  Then $\vn_i = \sum_{\beta=1}^m (\Vn_\beta/r_\beta^{2})/[\sum_{\beta=1}^m (1/r_\beta^{2})]$.  This fixes the $\vn_i$ on a narrow strip that contains the boundary $\Gamma$.  To fix the $\vn_i$ on the other nodes requires a velocity extension procedure, for which we use a fast-marching method, as in~\cite{Dunning2015}.

Given the $\vn_i$ on all nodes, the level set variables $\phi_i$ are updated according to (\ref{equ:advect-discrete}).  
The whole process -- inference of boundary points; sensitivity calculation; optimisation of $\lambda^{F}$ and $\lambda^{G}$; velocity extension; level set update -- is repeated until the algorithm converges to an optimal structure.  

Finally, note that in practice, it is convenient to do the velocity extension and the level set update only in a narrow band  close to the boundary.  This improves the efficiency of the method but it means that $\phi_i$ is given by the signed distance to the boundary only within the narrow band.  To correct for this effect, all of the $\phi_i$ variables are periodically reinitialised to be consistent with a signed distance function.  This reinitialisation uses the same fast-marching implementation used for the velocity extension.  

Note that given a set of boundary points $\XX_\alpha$, reinitialisation of the $\phi_i$ followed by a recalculation of the boundary point positions does in general lead to small changes in the boundary point positions (see appendix).  As the system gets close to convergence, this can have small but significant effects on the objective function, which acts as a weak source of numerical noise during optimisation.  Whilst this does not manifest itself as a significant concern in deterministic optimisation, it has implications for the accuracy of the stochastic method described below.

\section{Stochastic level set}
\label{sec:stochastic}

Having described the deterministic optimisation algorithm of~\cite{Dunning2015,Siva2016}, the next step is to introduce a stochastic component to this algorithm.  Several other stochastic level-set methods have been considered recently~\cite{Juan2006,Kasai2013,Zhou2016}, but differ from the approach proposed here in that the noise in this scheme is applied directly only on the boundary $\Gamma$, which ensures that the function $\phi$ retains its property as a signed-distance function under the stochastic time-evolution.

We first consider the optimisation problem for $F(\Omega)$ in the absence of any constraint on $G(\Omega)$.  In this case the deterministic algorithm corresponds to steepest descent for $F$, which converges to a local optimum.  By contrast, the corresponding stochastic algorithm does not converge to an optimum of $F$. Instead it converges to a steady state in which it explores a range of structures $\Omega$.  The algorithm is designed so that the probability $p_T$ that it visits a structure $\Omega$ within the steady state is proportional to a Gibbs-Boltzmann factor
\begin{equation}
p_T(\Omega) \propto {\rm e}^{-F(\Omega)/T}
\label{equ:gibbs}
\end{equation}
where $T$ is a noise intensity which would be identified as a temperature.
For $T\to0$ we see that the probability will concentrate close to the optimal structure $\Omega^*$, which minimises $F$.  

The advantage of the stochastic algorithm is that it can explore many different (non-optimal) designs.  In particular, for a non-convex optimisation problem and given a long enough time $t$, it should explore all the local optima, not just the one closest to the starting point (which would be found by a deterministic method).  Moreoever, the distribution of structures that the algorithm finds is controlled via (\ref{equ:gibbs}).
This means that the steady states of the algorithm at different temperatures are related, as 
\begin{equation}
\frac{p_{T_2}(\Omega)}{p_{T_1}(\Omega)} \propto {\rm e}^{F(\Omega)(1/T_1-1/T_2)}
\label{equ:rwt}
\end{equation}
and in particular, the marginal distribution of the objective function itself satifies
\begin{equation}
P_{T_2}(F) \propto {\rm e}^{F(1/T_1-1/T_2)} P_{T_1}(F)
\label{equ:rwt-F}
\end{equation}
Such relationships between distributions form the basis of simulated annealing and parallel tempering methods~\cite{frenkelsmit,Sugita1999,Rosta2009}, which have proven useful in exploring many non-convex optimisation problems.

Note however that these results are based on the proportionality relationship (\ref{equ:gibbs}).  More generally, we expect that the invariant measure of the stochastic process for the structures $\Omega$ is
\begin{equation}
\mathrm{d}p_T(\Omega) = \frac{1}{Z(T)} {\rm e}^{-F(\Omega)/T} \mathrm{d}p^0(\Omega)
\label{equ:POmega}
\end{equation}
where $Z(T)$ is a normalisation constant and $p^0$ is a reference measure for structures (independent of $T$). We do not have a precise characterisation of the measure $p^0$, but as long as $p^0$ is independent of $T$ then (\ref{equ:rwt},\ref{equ:rwt-F}) are valid, and can be used to check consistency between our algorithm and the asserted invariant measure (\ref{equ:POmega}).

\subsection{Stochastic motion of boundary points}

To describe our stochastic method, it is useful to cast (\ref{equ:advect-discrete}) as an equation of motion for the boundary points $\XX_\alpha$:
\begin{align}
\frac{\mathrm{d}}{\mathrm{d} t} \XX_\alpha(t) &= \Vn_\alpha(t) \bm{n}_\alpha(t) .
\label{equ:dXdet}
\end{align}
where $\bm{n}_\alpha$ is a unit vector in the direction of the (\rlj{inward-pointing}) normal to the boundary $\Gamma$ at point $\XX_\alpha$.
The stochastic element of the dynamics operates directly on the boundary points. To implement this, Eq.~(\ref{equ:dXdet}) is replaced by a \emph{stochastic differential equation} (SDE).  For the deterministic problem of minimisation of $F$ in the absence of any constraint, we write $\Vn_\alpha=-s^F_\alpha$, consistent with (\ref{equ:vnsf}).  Then a natural generalisation of (\ref{equ:dXdet}) is the SDE:
\begin{align}
\mathrm{d} \XX^\alpha_t &= -s^F_\alpha \bm{n}_\alpha \mathrm{d}t + \bm{n}_\alpha \sqrt{2T/\ell_\alpha} \circ \mathrm{d}W^\alpha_t
\label{equ:dXstoch}
\end{align}
The theory of such equations is discussed (for example) in~\cite{OksenBook}.  Roughly speaking, one may interpret $\mathrm{d}\XX^\alpha_t$ as a small increment in the boundary point position $\XX^\alpha$, associated with a small time increment $\mathrm{d}t$.  The increment $\mathrm{d} \XX^\alpha_t$ consists of a determinstic part $-s^F_\alpha \bm{n}_\alpha \mathrm{d}t$ that is proportional to $\mathrm{d}t$, and a random part $\bm{n}_\alpha \sqrt{2T/\ell_\alpha} \circ \mathrm{d}W^\alpha_t$.  The $\circ$ indicates that the SDE is written in the Stratonovich convention: the implications of this will be discussed below.
The increment $\mathrm{d}W^\alpha_t$ is a standard white noise (or Wiener process) associated with boundary point $\alpha$, which is independent of the noises on all other boundary points.

Inspection of (\ref{equ:dXstoch}) shows that the noise term acts in a direction perpendicular to the boundary (as might be expected); it is proportional to $\sqrt{T}$ which sets the noise strength.  The factor of $\sqrt{1/\ell_\alpha}$ might not be expected a priori -- it is necessary because consistency of the stochastic level-set method requires that the noise intensity is equal at all points on the boundary $\Gamma$, but the boundary points are not equally spaced along $\Gamma$ (recall Fig.~\ref{fig:boundary-integral}).  The idea is that a given boundary $\Gamma$ has many possible discretisations in terms of boundary points, but the resulting stochastic evolution should be independent of this discretisation.  Mathematically, this idea can be encapsulated as a reparameterisation invariance of the equations of motion, as we now discuss.

\subsection{Evolution of continuous boundaries, and reparameterisation invariance }
\label{subsec:stoch-boundary}

It is useful to consider a process by which a continuous curve evolves in time, and to interpret (\ref{equ:dXstoch}) as a discretised approximation of this process.  For the continuous curve we write a generalisation of (\ref{equ:vnsf}), as
\begin{align}
\mathrm{d} \XX_t(u) &= -s^F(u) \bm{n}(u) \mathrm{d}t + \bm{n}(u) \sqrt{2T/\ell'(u)} \circ \mathrm{d}W_t(u)
\label{equ:dXu-stoch}
\end{align}
where $W_t(u)$ a random Brownian noise with equal intensity at each point $u$. (Technically, the quadratic variation  of this process satisfies $\mathrm{d} \langle \int_{u_1}^{u_2} W_t(u) \mathrm{d}u \rangle = (u_2-u_1)$.)  To derive (\ref{equ:dXstoch}) as a discretised version of (\ref{equ:dXu-stoch}), we identify $\XX^\alpha$ as the centre of mass of a small segment of the boundary: $\XX_t^\alpha = (1/\ell_\alpha) \int_{u_1}^{u_2} \XX_t(u) \mathrm{d}\ell(u)$.  Assuming (see below) that the segment is small enough that $\bm{n}(u)$ and $\ell'(u)$ and $s^F(u)$ can be replaced by their mean values within the integral, this yields (\ref{equ:dXstoch}).

Within this continuous setting, the idea that the evolution of the curve should be independent of its discretisation in terms of boundary points has a mathematical statement in terms of reparameterisation of the coordinate $u$.  Consider a closed curve described by the internal co-ordinate $u\in[0,1]$, so $\XX : [0,1]\to \Omega_d$ is a smooth function with $\XX(u)$ a point on the curve.
Now consider 
a continuous monotonically-increasing function $f : [0,1]\to[0,1]$ with $f(0)=0$ and $f(1)=1$.  Define $\tilde\XX : [0,1]\to\Omega_d$ by $\tilde\XX(u)=\XX(f(u))$.  It should be clear that $\XX$ and $\tilde \XX$ are different representations of the same curve.  

In order that the evolution of a curve $\Gamma$ does not depend on its parameterisation in terms of boundary points, we require that (\ref{equ:dXu-stoch}) evolves the functions $\XX$ and $\tilde\XX$ in the same way.  This may be verified by direct substitution, as long as the noise prefactor $\sqrt{1/\ell'(u)}$ is included.  This is the reason for including the factor in $\sqrt{1/\ell_\alpha}$ that multiplies the noise in (\ref{equ:dXstoch}).

We have assumed throughout that the boundary $\Gamma$ is smooth enough and the normal vector $\bm{n}(u)$ can be defined at every point $\bm{X}(u)$.  Even for deterministic optimisation, this assumption requires some care, if the optimal shape has a boundary that includes kinks.  For stochastic optimisation, there is an additional factor, which is that the noise tends to roughen the curve and the normal vector $\bm{n}(u)$ may not be defined.  In our numerical scheme, the boundary is discretised and a WENO estimate of $\nabla\phi$ can always be used to define a local normal, so potential problems with rough boundaries do not appear.  For a detailed mathematical analysis, some regularisation is required to ensure existence and uniqueness of the SDE (\ref{equ:dXu-stoch}), but we defer these issues to a later publication.%

\subsection{Invariant measure}

Our assertion is that the invariant measure for the stochastic process described in Sec.~\ref{subsec:stoch-boundary} is proportional to (\ref{equ:gibbs}).  We do not have a rigorous proof of this claim, although we will demonstrate below that our numerical method is consistent with (\ref{equ:rwt-F}).  As a plausibility argument for our scheme, we note that if $\bm{n}_\alpha s^F_\alpha$ in (\ref{equ:dXstoch}) is equal to $(\partial F/\partial \bm{X}_\alpha)\ell_\alpha^{-1}$ then (\ref{equ:dXstoch}) has the form
\begin{equation}
\mathrm{d}\bm{X}^\alpha_t = (\partial F/\partial \bm{X}_\alpha) \sigma^2 \mathrm{d}t + \sigma \sqrt{2T} \circ \mathrm{d}W_t^\alpha
\end{equation}
with $\sigma=1/\sqrt{\ell_\alpha}$.  It is well-known~\cite{OksenBook} that the invariant measure for this equation is of the form (\ref{equ:gibbs}).  Moreover, spatially discretising (\ref{equ:sensF}) and identifying $\eps z(u)$ as the boundary point displacement $\bm{n}_\alpha \cdot \delta \bm{X}_\alpha$ yields
\begin{equation}
F(\Omega_\eps) - F(\Omega) = \sum_\alpha s^F_\alpha \ell_\alpha \bm{n}_\alpha \cdot \delta \bm{X}_\alpha + O(\delta \bm{X}^2)
\end{equation}
indicating that $\ell_\alpha \bm{n}_\alpha s^F_\alpha$ does indeed correspond to the derivative $(\partial F/\partial \bm{X}_\alpha)$, as required.  Refining this argument into a more rigorous proof is an interesting direction for future work.

In the following, we show how the process (\ref{equ:dXstoch}) can be implemented within the level-set method described in Sec.~\ref{sec:level-set}.  We assume that (\ref{equ:rwt}) holds theoretically and we investigate the extent to which it holds numerically.

\begin{figure}[t]
\centering
\includegraphics[width=8cm]{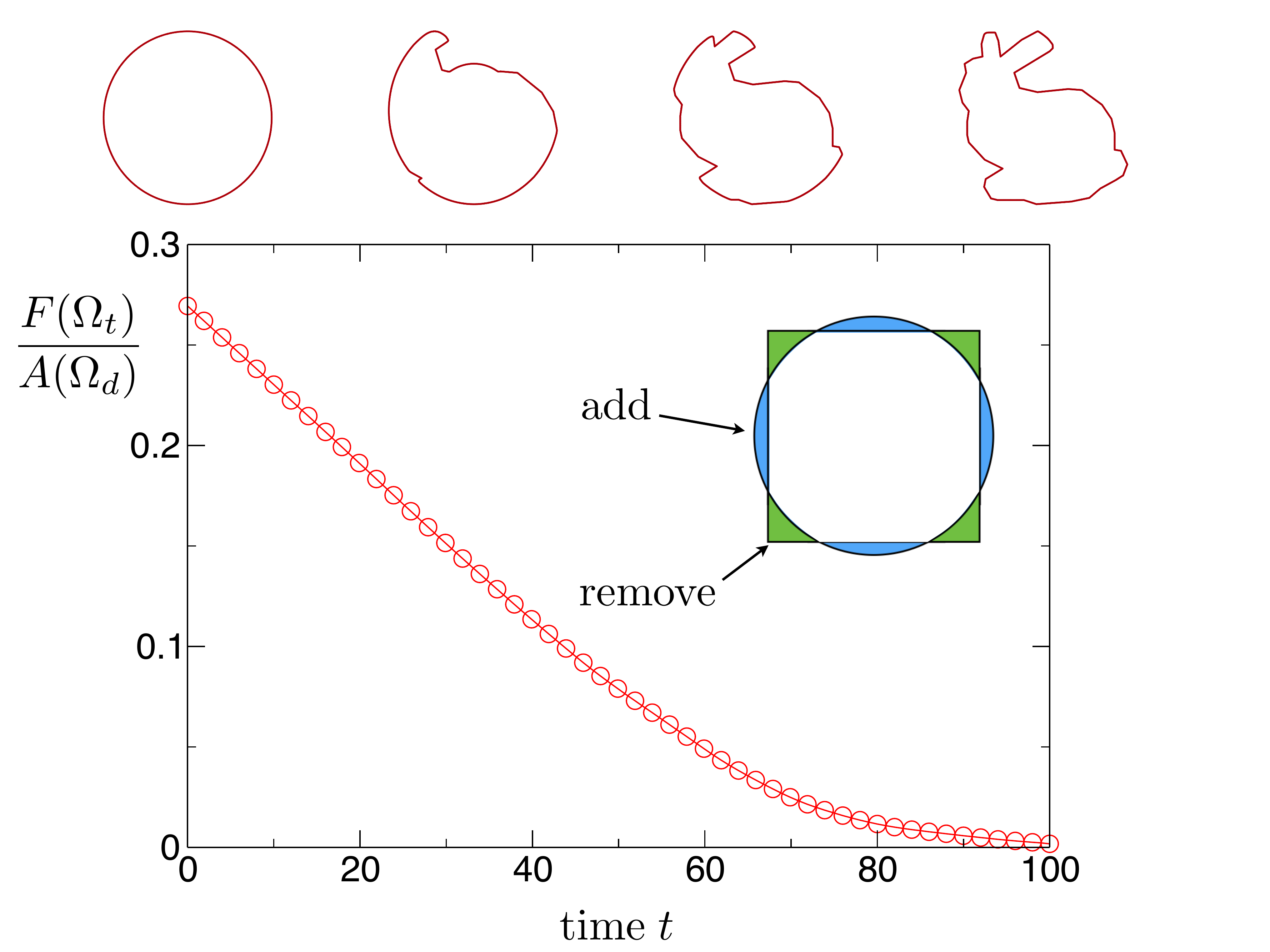}
\caption[Demonstration of shape matching by the level-set method (deterministic)]{\label{fig:bunny1}%
Shape matching by the level-set method (without any stochastic component). The target shape is a discretised representation of the Stanford bunny (top right); the initial shape is a circle (top left).
The main figure shows a time series of the (normalised) objective function $F(\Omega_t)/A(\Omega_{\rm d})$ with $F$ given by (\ref{equ:F-match}) and $A(\Omega_{\rm d})$ the total area of the design domain.
Representative shapes obtained during this convergence procedure are also shown (above). The inset illustrates the definition of the objective and sensitivity functions, for the simpler case of a circular target shape and and a square initial shape $\Omega$.  Blue shading indicates regions where $\phi_i^{\mathrm{target}} > \phi_i$ and \rlj{$s^F=1$}, so \rlj{inward} motion of the boundary \rlj{increases} $F$.  Similarly green regions have $\phi_i^{\mathrm{target}} < \phi_i$ and \rlj{$s^F=-1$}.  The normalised objective function $F(\Omega)/A(\Omega_{\rm d})$ is given by the sum of the blue and green areas, as a fraction of the total area of the design domain $\Omega_{\rm d}$.}
\end{figure}

\begin{figure*}[tb]
\centering
\includegraphics[width=0.85\linewidth]{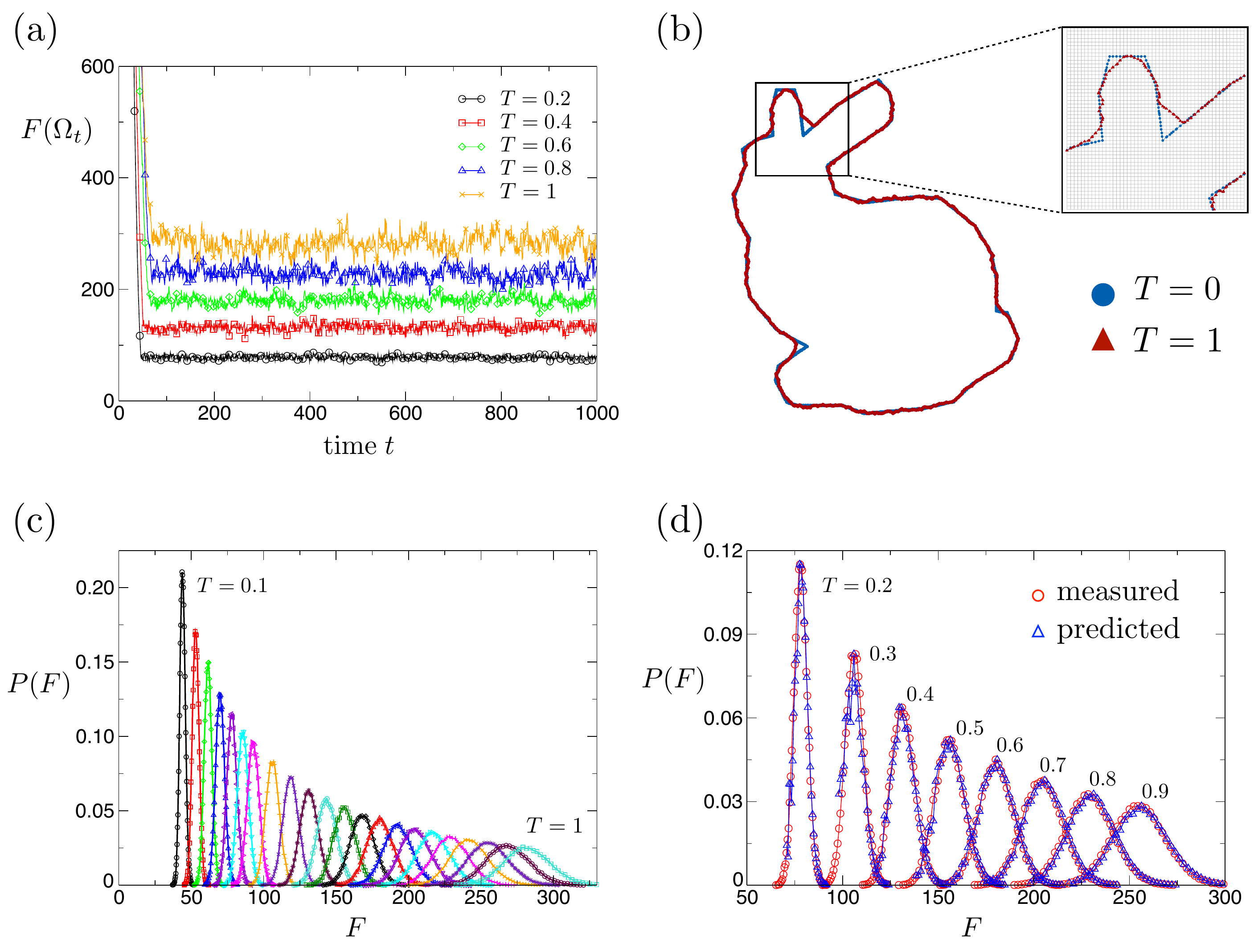}
\caption[Demonstration of shape matching with finite noise.]
{\label{fig:bunny2} {Shape matching with finite noise strength $T$: (a) Time series of the area mismatch (\ref{equ:F-match}) for several noise strengths (temperatures). Raising temperature leads to increasingly sub-optimal designs and larger fluctuations. (b) A representative shape (red) from the steady state at temperature $T=1$, compared with the target shape (red). Mismatch with the target structure is most pronounced in regions of high curvature. (c) Histograms (normalised as probability densities) of the objective function, obtained from long time series in the steady state of the system. Data are shown for seven temperatures equally spaced between $T=0.1$ and $T=0.25$, with the remaining temperatures equally spaced between $T=0.25$ and $T=1.0$. (d) The measured probability densities agree well with the predictions of Eq.~(\ref{equ:rwt-F}): we show measured $p_T(F)$ for various temperatures, and the corresponding predictions using Eq.~(\ref{equ:rwt-F}) with $T_1=T$ and $T_2=T+\Delta T$.  The temperature increment $\Delta T$ is such that the distributions at $T_1,T_2$ are adjacent in (c).}}
\end{figure*}

\subsection{Stochastic dynamics with a finite time step}

To implement the stochastic evolution (\ref{equ:dXstoch}) within the level-set method of Sec.~\ref{sec:level-set}, we require a generalisation of (\ref{equ:advect-discrete}).  To achieve this, integrate (\ref{equ:dXstoch})  over a small time interval $\Delta t$ and identify the average boundary point velocity in the normal direction as $\overline{\Vn_\alpha}(t)=(\Delta t)^{-1}\int_{s=t}^{t+\Delta t} \bm{n}(s) \cdot \mathrm{d}\XX^\alpha_s$. Hence, keeping terms up to first-order in $\Delta t$:
\begin{equation}
\overline{\Vn_\alpha}(t)\Delta t = -s^F_\alpha(t) \Delta t + \sqrt{\frac{2T\Delta t}{\ell_\alpha(t+\Delta t/2)}} \xi 
\label{equ:vnbar-strat}
\end{equation}
where $\xi$ is a standard normal-distributed random number.  The first term on the right hand side is the standard deterministic increment corresponding to (\ref{equ:dXdet}).  The second term is a random increment (with zero mean).  Note however that the length $\ell_\alpha$ in this second term is evaluated at time $t+\Delta t/2$, which corresponds to the midpoint of the time interval (this is the meaning of the Stratonovich product that was denoted by $\circ$ in (\ref{equ:dXstoch})).  To arrive at an equation that can be used in a (first-order) numerical scheme, it is necessary to calculate $\overline{\Vn_\alpha}(t)$ in terms of quantities that are evaluated only at time $t$.  This can be achieved by using Ito's formula~\cite{OksenBook}, which yields 
\begin{equation}
\overline{\Vn_\alpha}(t)\Delta t = -s^F_\alpha(t) \Delta t + \sqrt{\frac{2T\Delta t}{\ell_\alpha(t)}} \xi   -\frac{T \kappa_\alpha}{2\ell_\alpha} \Delta t .
\label{equ:vnbar}
\end{equation}
where we again keep terms up to first-order in $\Delta t$, and $\kappa_\alpha$ is the (signed) curvature of the boundary at point $\alpha$ (see Appendix).

The derivation of (\ref{equ:vnbar}) from (\ref{equ:dXstoch}) is straightforward within the framework of stochastic calculus.  We omit technical details and provide a short argument to justify it: for a boundary point increment $\mathrm{d}\bm{X}_\alpha$, the change in the length of the boundary segment is $\mathrm{d}\ell_\alpha = \bm{n}_\alpha \cdot (\ell_\alpha\kappa_\alpha \circ \mathrm{d}\bm{X}_\alpha)$.  (Roughly speaking this corresponds to the chain rule, with $\mathrm{d}\ell_\alpha/\mathrm{d}\bm{X}_\alpha = \bm{n}_\alpha\ell_\alpha\kappa_\alpha$, as discussed in the Appendix.)  Ito's formula~\cite{OksenBook} states that for an SDE of the form $\mathrm{d}x_t = f(x_t) \mathrm{d}t + \sigma(x_t) \circ \mathrm{d}W_t$, the appropriate first-order discretisation is $\Delta x = f(x_t) \Delta t + \sigma(x_t) \xi \sqrt{\Delta t} + \frac12 \sigma'(x_t) \sigma(x_t) \Delta t$.  In this case we have from (\ref{equ:dXstoch}) that $\sigma=\bm{n}_\alpha\sqrt{2T/\ell_\alpha}$: the analogue of $\sigma'$ is $(d\sigma/d\ell_\alpha)\cdot (d\ell_\alpha/d{\bm X}_\alpha)=(-\kappa_\alpha/2)\sqrt{2T/\ell_\alpha}$, which yields (\ref{equ:vnbar}).

\subsection{Implementation in level-set method, and incorporation of constraints}

The stochastic velocity (\ref{equ:vnbar}) is straightforwardly incorporated into the deterministic level set method described in Sec.~\ref{sec:level-set}.  This yields our stochastic level set optimisation method.  The only modification  of the deterministic algorithm is that noise-dependent terms are added to (\ref{equ:vn-lambda-disc}).  We consider the case of optimisation subject to the constraint (\ref{equ:GG}).

In the stochastic method, the parameters $\lambda^F,\lambda^G$ in (\ref{equ:vn-lambda-disc}) are calculated exactly as in the deterministic method.  The simplest approach is then to identify $\lambda^F\to-\Delta t$ and add the stochastic terms in (\ref{equ:vnbar}) to the boundary increments.  However, since the first stochastic term is proportional to $\sqrt{\Delta t}$, this can lead to large boundary point displacements that violate the CFL condition.  Hence, given a deterministic time step $\Delta t = -\lambda^F$, we calculate a typical stochastic increment $\Delta x_{\rm typ} = \sqrt{2T\Delta t}$.  If $\Delta x_{\rm typ} > d_{\rm CFL}/2$, the deterministic parameters $\lambda^F,\lambda^G$ are rescaled by a factor $d_{\rm CFL}/2\Delta x_{\rm typ}$.  That is 
\begin{equation}
\begin{split}
\lambda^F_{\rm stoch} &= \lambda^F\, {\rm min}(1,d_{\rm CFL}/2\Delta x_{\rm typ}) \\
\lambda^G_{\rm stoch} &= \lambda^G\, {\rm min}(1,d_{\rm CFL}/2\Delta x_{\rm typ}) .
\end{split}
\end{equation}
This construction amounts to fixing a maximal time step that ensures that the CFL condition is obeyed even in the presence of the noise, as long as none of the $\ell_\alpha$ parameters are too small (compared to unity).  (We also tested a method where the $\Delta x_{\rm typ}$ is adjusted to account for the possibilities that some $\ell_\alpha$s are very small, but this led to a less efficient method and had very little effect on the results.)
The time step is $-\lambda^F_{\rm stoch}$ and the boundary point increment is finally
\begin{equation}
\overline{\Vn_\alpha}\Delta t = s^F_\alpha \lambda^F_{\rm stoch} + \sqrt{\frac{2T|\lambda^F_{\rm stoch}|}{\ell_\alpha}} \xi   +\frac{T \kappa_\alpha}{2\ell_\alpha} \lambda^F_{\rm stoch} + \lambda^G_{\rm stoch} s^G_\alpha
\label{equ:vnbar-stoch-lambda}
\end{equation}
This equation replaces (\ref{equ:vn-lambda-disc}) within our stochastic level set method.  Once the boundary point velocities $\overline{\Vn_\alpha}$ have been calculated in this way, the rest of the method follows exactly the deterministic case: the velocities on the nodes are calculated by interpolation and fast-marching, and the level set variables are updated using (\ref{equ:advect-discrete}).

\section{Results}
\label{sec:results}

{In the following we consider several examples of the stochastic level set method, with increasing levels of complexity}

\begin{figure}[tb]
\centering
\includegraphics[width=7.5cm]{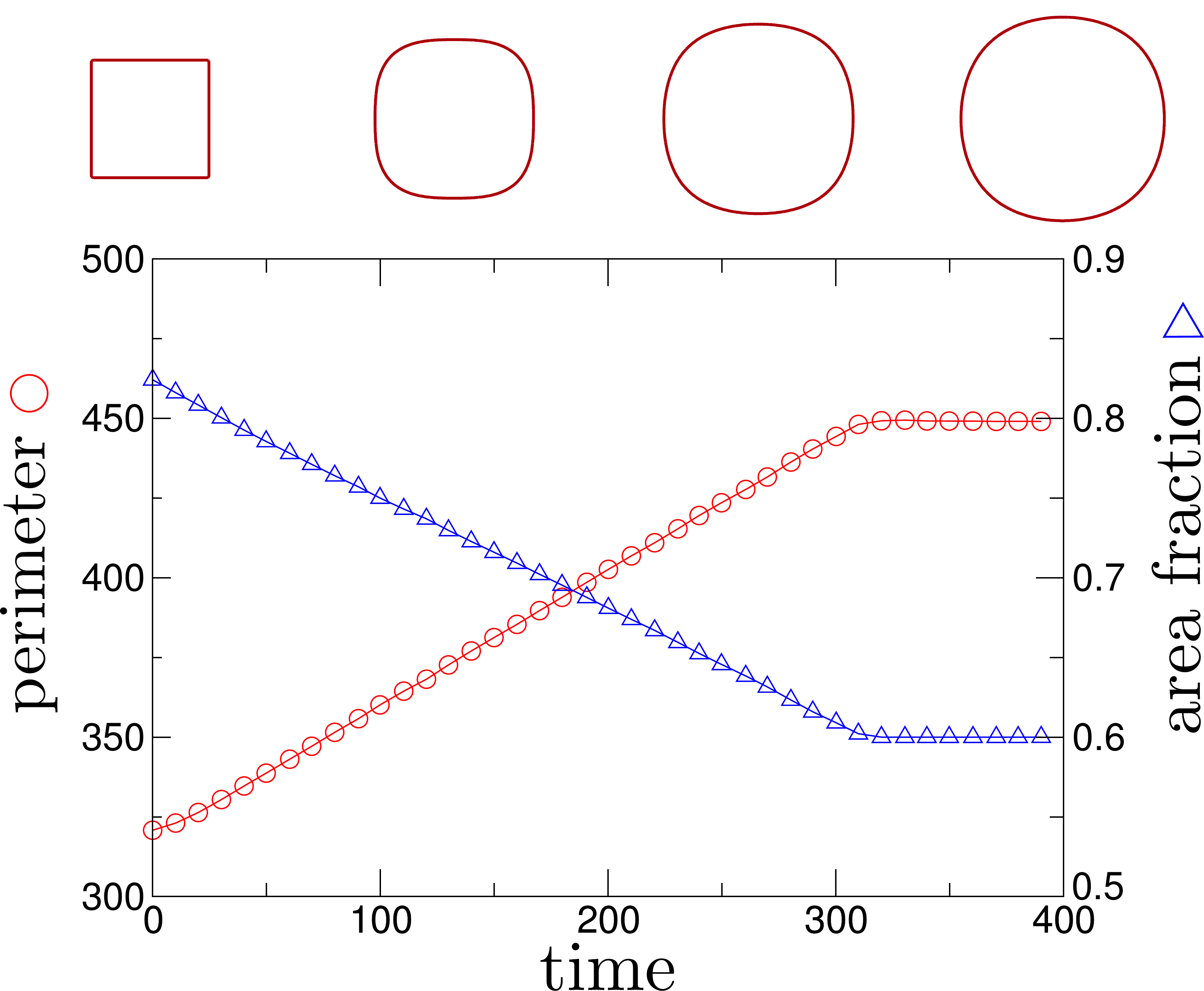}
\caption[Minimisation of shape perimeter at constant area (deterministic)]
{\label{fig:perimeter1} Time series for deterministic minimisation of shape perimeter (circles, left axis), subject to a constraint that the area fraction \emph{outside} $\Omega$ must be at most 60\% of the total area (triangles, right axis).  The top panel shows the time-dependence of $\Omega$, starting from a square initial shape, which violates the constraint.  The shape $\Omega$ increases in size (to satisfy the constraint) and evolves towards a circle (to minimise the perimeter). The domain $\Omega_{\rm d}$ is a square grid of size $200 \times 200$.}
\end{figure}

\begin{figure*}[tb]
\centering
\includegraphics[width=0.9\linewidth]{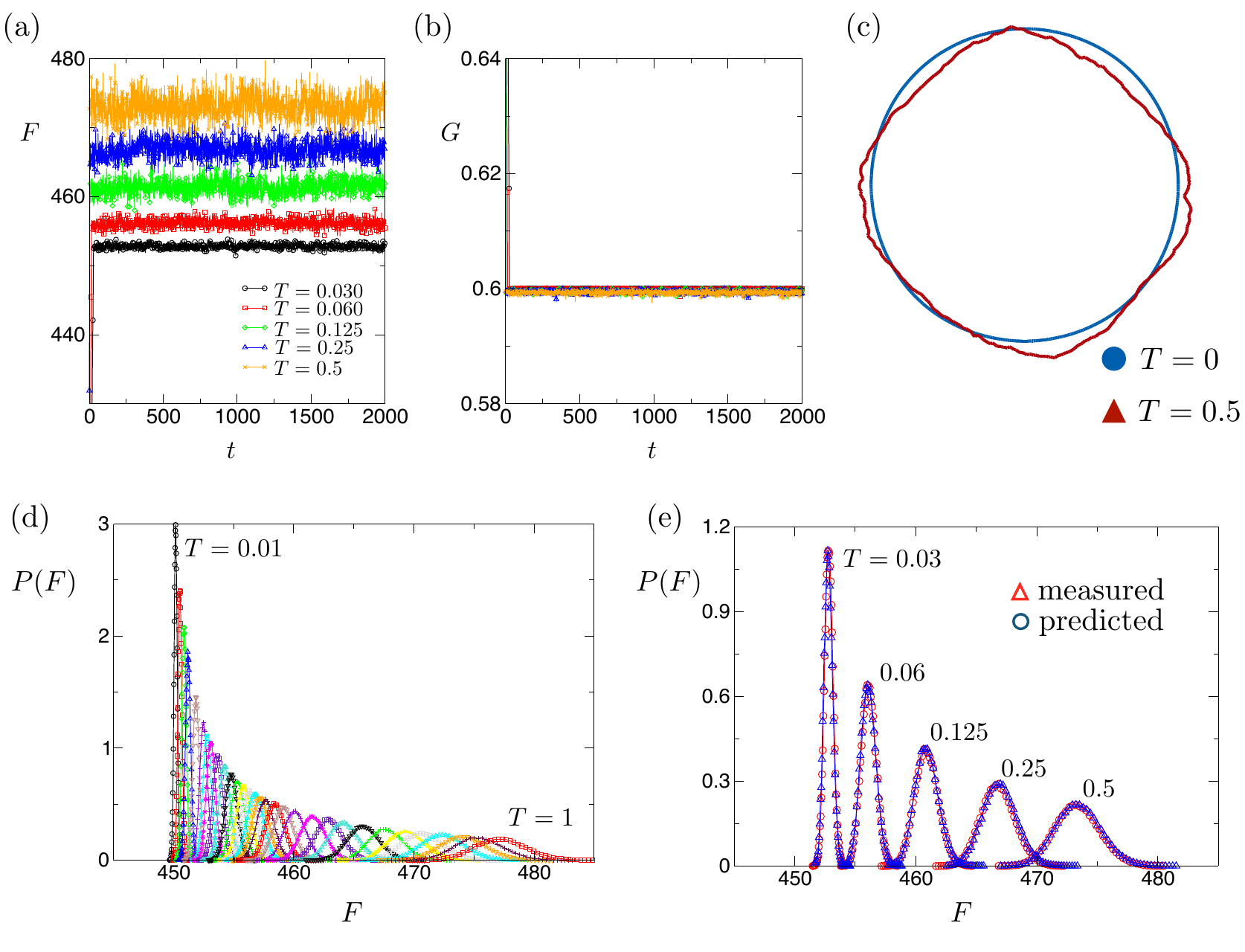}
\caption[Minimisation of shape perimeter at constant area (with finite noise)]
{\label{fig:perimeter2} Stochastic optimisation of perimeter with a constraint on the enclosed area.  (a,b)~Time series for the objective function $F$ (perimeter) and the constraint function $G$ (area). The legend in (a) applies to both panels.  (c)~The zero-temperature optimum (a circle), compared with a representative sample from a stochastic calculation at $T=0.5$.  (d)~Histograms of the objective function, taken from the steady state of the stochastic process, at various temperatures.  (e)~Measured probability densities, compared with the predictions of Eq.~(\ref{equ:rwt-F}). The predictions use Eq.~(\ref{equ:rwt-F}) with $T_1=T$ and $T_2=T+\Delta T$, as in Fig.~\ref{fig:bunny2}.  The agreement is good.
}
\end{figure*}

\subsection{Shape matching}
\label{sec:shape-matching}

We first consider a simple geometric optimisation problem.  The idea, illustrated in Fig.~\ref{fig:bunny1} is that the domain $\Omega$ should match a predefined reference shape $\Omega_{\rm target}$, in this case a discretised two-dimensional representation of the Stanford bunny~\cite{bunny}.  The objective function is  
\begin{equation}
    F(\Omega) = \sum_{i} |A_i^{\mathrm{target}} - A_i(\Omega)|
    \label{equ:F-match}
\end{equation}
where the sum runs over the cells of the grid, $A_i(\Omega)$ is the overlap area between the grid cell $i$ and $\Omega$, and similarly $A_i^{\mathrm{target}}$ is the overlap area between element $i$ and $\Omega_{\rm target}$.

To calculate sensitivities, note that if $A$ is the area of $\Omega$ then \rlj{$s^A_\alpha=-1$} for all boundary points $\alpha$.  For the objective function (\ref{equ:F-match}), this requires \rlj{$s^F_\alpha=1$} if boundary point $\alpha$ is inside $\Omega_{\rm target}$ and \rlj{$s^F_\alpha=-1$} otherwise.  To estimate this sensitivity based on local information, we calculate for each node $i$ its signed distance $\phi_i^{\rm target}$ from the boundary of the target shape.  For each boundary point $\alpha$, we estimate its signed distance from the boundary of the target as $\Phi_\alpha^{\rm target}$, which is interpolated from the $\phi_i^{\rm target}$ values on the four nearest nodes.  We also interpolate  a value $\Phi_\alpha^{\rm trial}$ by applying the same method to the nearest values of $\phi_i$: since the boundary point is on the zero contour of $\phi$ we expect $|\Phi_\alpha^{\rm trial}|\ll 1$ for all $\alpha$.  The sensitivity is then estimated as
\begin{equation}
    s^F_\alpha = \mathrm{sign}( \Phi_\alpha^{\mathrm{target}} - \Phi_\alpha^{\mathrm{trial}} )
    \label{equ:sens-match}
\end{equation}
Since  $\Phi_\alpha^{\mathrm{trial}}$ is small in magnitude, its inclusion in (\ref{equ:sens-match}) has a small effect when the current and target shapes are different from each other, but ensures that the deterministic method converges (exactly) to $\Omega=\Omega^{\rm target}$.

Based on these sensitivities, Fig.~\ref{fig:bunny1} shows the deterministic optimisation algorithm in operation.  An initially circular structure evolves in time until it matches exactly the target shape.  The CFL constraint is $d_{\rm CFL} = 0.1$ and the domain is partitioned into a grid with $200\times200$ elements, with each element having size $1$.  The initial shape is a circle of radius $50$.

Fig.~\ref{fig:bunny2} shows results for the stochastic algorithm, applied to the same problem.  Panel (a) shows that the objective function does not converge to its optimal value ($F=0$).  Instead the system converges to a steady state in which $F$ fluctuates around a non-zero mean value.  As the noise strength increases, both the mean value of $F$ and its fluctuations increase, as the algorithm explores a range of non-optimal designs.  Panel (b) shows a representative non-optimal shape.  Panel (c) shows distributions (histograms) of the objective function, for different values of the noise.  Finally in panel (d) we test the prediction of (\ref{equ:rwt-F}), that distributions of the objective function at different noise strengths can be related to each other.  That is, for various temperatures $T$, the distribution $P(F)$ can be predicted based on its distribution at a different temperature $T+\Delta T$.  (This prediction is possible only for small $\Delta T$ since otherwise the exponential factor in (\ref{equ:rwt-F}) leads to large statistical errors.)

The results shown in Fig.~\ref{fig:bunny2}(d) represent strong evidence that the system has converged to a steady state in which the predictions of (\ref{equ:rwt-F}) are valid.  The stochastic algorithm was designed in order to obtain a steady state that satisfies (\ref{equ:gibbs}), and the numerical agreement with (\ref{equ:rwt-F}) is a stringent test of this criterion, which indicates that the theoretical analysis and numerical implementation presented here are self-consistent.  

Moreoever, the nature of the non-optimal shapes shown in Fig.~\ref{fig:bunny2}(b) reveal some information about the null measure $p_0$ in (\ref{equ:POmega}): of the shapes $\Omega$ for which $F(\Omega)$ is not equal to its optimal value, the method seems to sample preferentially those shapes with low curvature.  In this sense the numerical noise seems to act to smooth the boundary $\Gamma$.  A detailed analysis of this effect will require a deeper understanding of the stochastic dynamics of the boundary, which is an interesting direction for future work.

\subsection{Perimeter minimisation with area constraint}
\label{sec:perimeter}

As a simple optimisation problem that includes a constraint, we next consider minimisation of the perimeter of $\Omega$, subject to a constraint on its total area. Given that $F(\Omega)$ is the perimeter of $\Omega$, the sensitivity $s^F$ is  is given by the (signed) Euclidean curvature $\kappa$ of the boundary $\Gamma$ (see Appendix).  The constraint function  $G$ is the difference between the total domain size and the area of $\Omega$: that is, $G(\Omega)=A(\Omega_{\rm d})-A(\Omega)$.  (This ensures that the \emph{lower} bound on the area of $\Omega$ can be cast as a constraint in the form (\ref{equ:GG}).)  The associated sensitivity is \rlj{$s^G=1$}.

Estimation of the curvature $\kappa$ is non-trivial for the discretised boundaries considered here~\cite{Calabi1998}.  Following numerical tests of several different methods, we use a scheme that combines accuracy and simplicity, by calculating curvatures using an explicit finite-difference sensitivity calculation.  Details of this procedure, and its associated errors are discussed in the Appendix.

Results for deterministic optimisation are shown in Fig.~\ref{fig:perimeter1}.  The grid is of size $200\times 200$.  An initially square structure $\Omega$ evolves into a circle whose area satisfies the constraint. In this case $G\leq 0.6A(\Omega_{\rm d})$, so the area of $\Omega$ must be at least $40\%$ of the domain $\Omega_{\rm d}$.  

Fig.~\ref{fig:perimeter2} shows results for the stochastic method, which are comparable with Fig.~\ref{fig:bunny2}.  Fig.~\ref{fig:perimeter2}(b) shows that the stochastic algorithm still satisfies the constraint (up to some numerical uncertainties).  Fig.~\ref{fig:perimeter2}(e) shows that the results are again consistent with the algorithm sampling a distribution of shapes consistent with (\ref{equ:gibbs},\ref{equ:POmega}).

However, the example shape shown in Fig.~\ref{fig:perimeter2}(c) for $T=0.5$ reveals that this numerical method is affected by discretisation from the underlying grid.  In particular, the shape $\Omega$ shown in that figure is not circular, but is elongated along the lattice axes, forming a kind of diamond shape.  We find that the shapes $\Omega$ found for $T>0$ consistently have this property (data not shown).  We have investigated the reason for this effect, which we attribute to uncertainties in our numerical estimates of the sensitivity parameters $s_\alpha$, particularly the sensitivity of the perimeter, which is the curvature.  These numerical issues are discussed in the Appendix.  In terms of the general method, our conclusion is that the stochastic level-set method relies on accurate sensitivity estimates, which may be difficult to obtain in practice.  However, we believe that the method itself is valid.

\subsection{Non-convex shape matching problem with perimeter constraint}

\begin{figure}[tb]
\centering
\includegraphics[width=8cm]{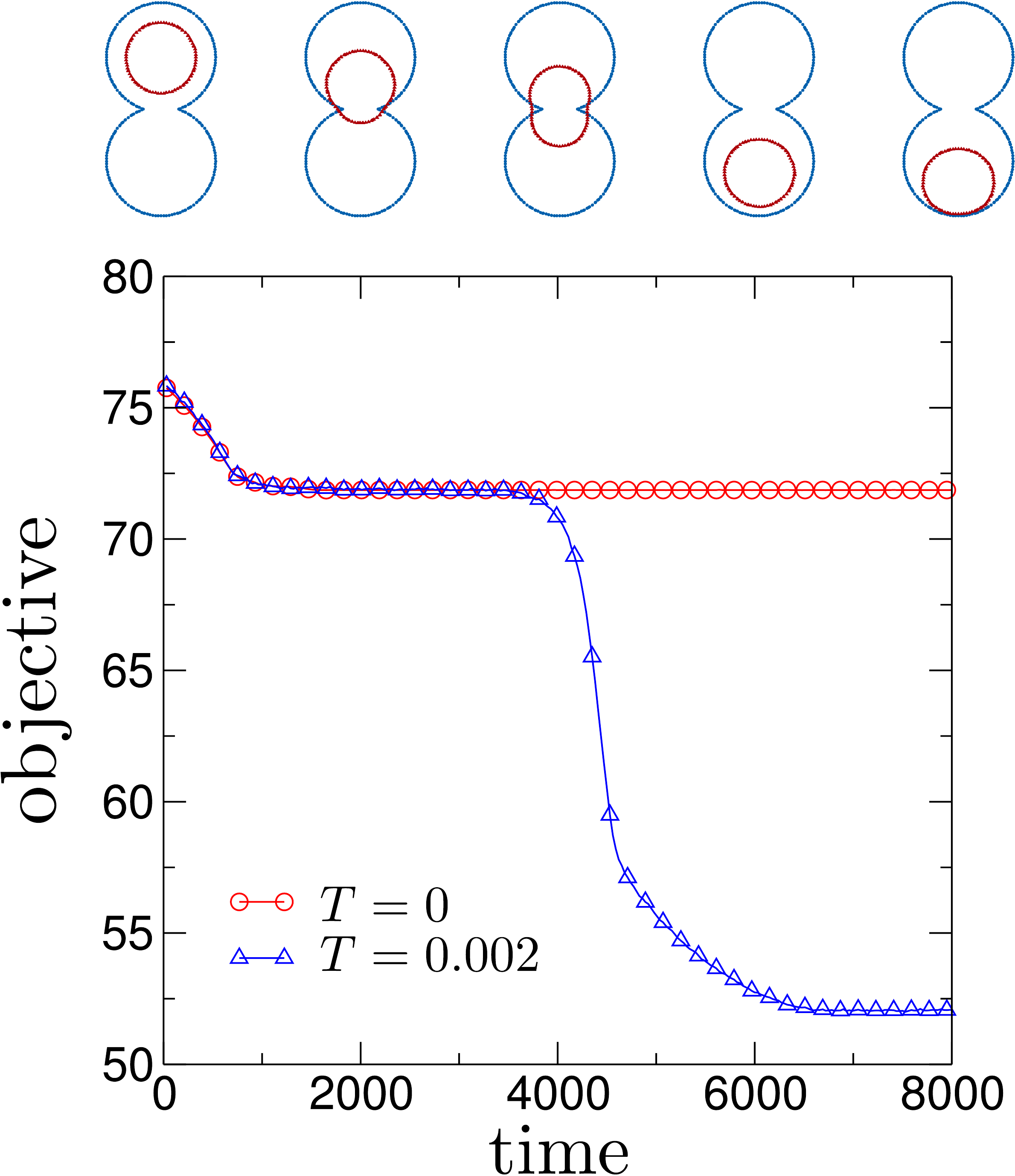}
\caption[Non-convex optimisation problem]
{\label{fig:dumb} Non-convex optimisation problem.  (Top row) snapshots from an optimisation trajectory in which the perimeter of the red shape $\Omega$ and its height within the design domain are both being minimised, subject to a constraint on the mismatch of the red shape with the blue ``dumbbell'.  The global optimum is shown in the rightmost panel while the second panel illustrates an additional local optimum.  (Main panel) Time series of the objective function for deterministic optimisation ($T=0$) and stochastic optimisation ($T=0.002$).  Only in the stochastic case does the system escape from the local optimum and converge to the global one.  The mesh size is $100\times 100$, the dumbell consists of two circles of radius $20$ whose centres are separated by $B=38$. The parameter $\gamma=0.65$.}
\end{figure}

So far, we have focussed on very simple optimisation problems, as proof of principle for the method.  However, a central motivation for the development of a stochastic method is that the noise forces can allow the system to visit multiple locally-optimal designs in a non-convex optimisation problem.  As a  simple example of such a problem, we consider the problem shown in Fig.~\ref{fig:dumb}.  

The constraint function in this case is related to the overlap between the shape $\Omega$ and the dumbbell (or hourglass) target shape shown in Fig.~\ref{fig:dumb}.  By analogy with (\ref{equ:F-match}), we define
\begin{equation}
	G = \sum_{i} |A_i^{\mathrm{target}} - A_i(\Omega)|
\end{equation}
where the sum runs over all cells of the underlying grid.  The constraint is that $G\leq 0.2A(\Omega_d)$: that is, the mismatch between the shape $\Omega$ and the target (dumbbell) shape can be at most $20\%$ of the total design domain. In practical terms, this means that when $\Omega$ is contained within the dumbbell, the area of the dumbbell that is outside $\Omega$ must be at most $0.2A(\Omega_d)$. 
The dumbbell is given by the union of two circles of radius 20.  In Cartesian coordinates, the size of the design domain is $L_x\times L_y$ and the centres of the circles are at $(L_x/2,(L_y\pm B)/2)$ with $B=38$.

The objective function in this case is a weighted perimeter $F(\Omega)=M(\Omega)$ with
\begin{equation}
M(\Omega)= \int_\Gamma m(\bm{X}(u)) \mathrm{d}\ell(u)
\label{equ:F-dumb}
\end{equation}
If $m(\bm{X})=1$ for all $\bm{X}$ then $M$ is simply the perimeter of $\Omega$.  The idea is that $m(\bm{X})$ is the weight (per unit length) of a boundary segment at position $\bm{X}$.  We write $\bm{X}=(X,Y)$ in Cartesian coordinates and $m(\bm{X})$ depends only on $Y$.  It is given by $m=1$ for positions above the centre of the upper lobe of the dumbbell, and $m=\gamma$ for positions below the centre of the lower lobe.  In between, $m$ varies linearly with $Y$, so that
\begin{equation}
m(X,Y) = 
\begin{cases} 1, & \; Y > (L_y + B)/2 \\
\gamma, & \; Y < (L_y - B)/2 \\
\frac{1+\gamma}{2} + \frac{1-\gamma}{2}(Y-L_y/2), & \; {\rm otherwise}  .
\end{cases}
\end{equation}
The parameter $\gamma=0.65$. 
Physically, the objective function is small when the boundary $\Gamma$ is located in the lower lobe of the dumbbell, and larger when it is in the upper lobe.  The optimal design is a circle located inside the dumbbell, below the centre of the lower lobe.

\begin{figure*}[t]
\centering
\includegraphics[width=\linewidth]{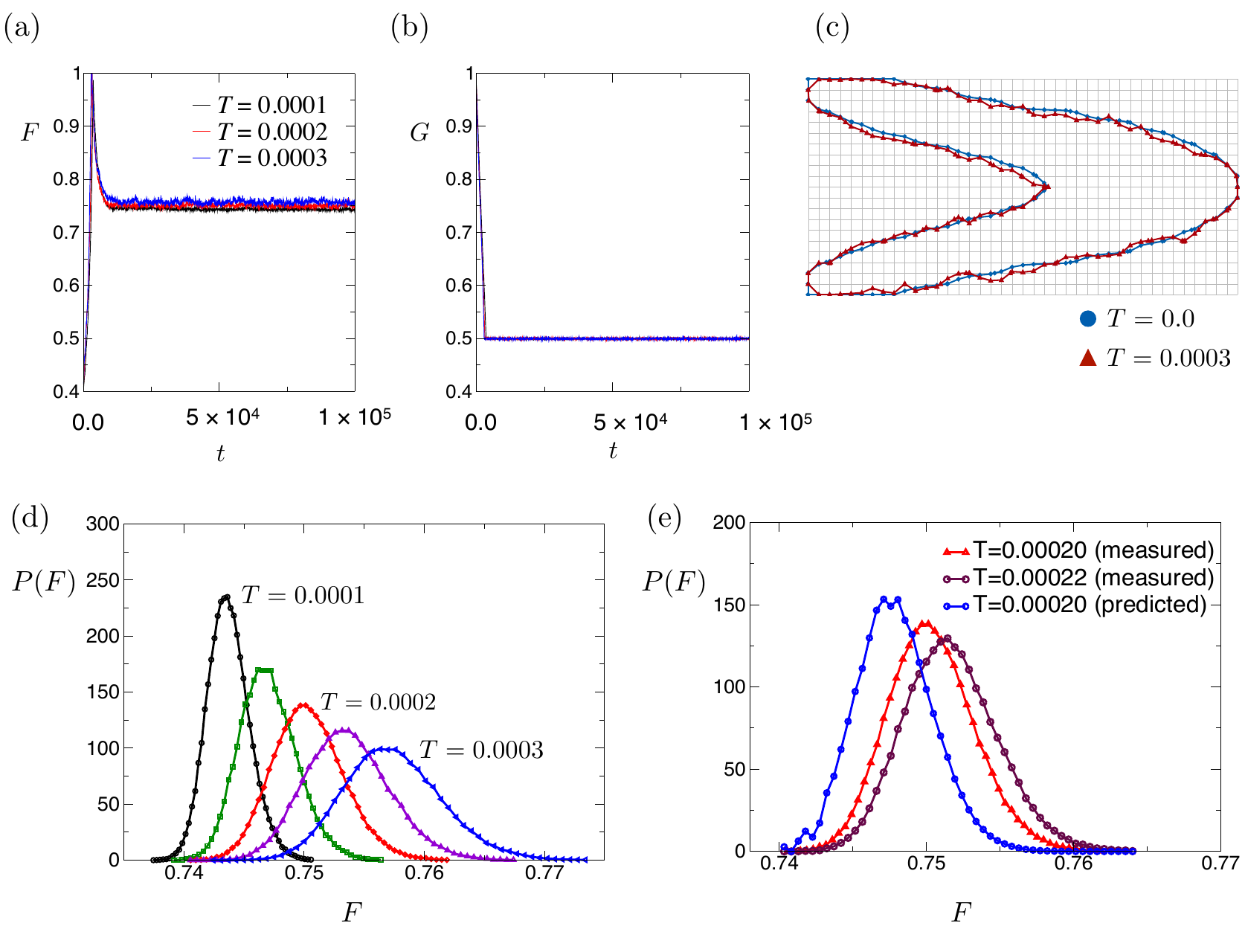}
\caption[Optimisation of compliance (with noise)]
{\label{fig:compliance} Stochastic optimisation of compliance.  (a,b) Time series for the objective function (strain energy) and constraint function (material area) for the model problem described in the text.  (c) The (local) minimum found by deterministic optimisation $(T=0)$ and a representative structure at $T=0.0003$.  (d) Histograms for the objective function for different noise strengths $T$.  (e) Test of the reweighting formula (\ref{equ:rwt-F}).  The results are not  consistent with (\ref{equ:rwt-F}): we show measured distributions at $T_1=0.00020$ and $T_2=0.00022$.  The distribution at $T_2$ is used together in (\ref{equ:rwt-F}) to arrive at a prediction for the distribution at $T_1$, but this prediction is not accurate.  We attribute this effect to numerical errors associated with discretisation and sensitivity estimation, as discussed in the main text }
\end{figure*}

The sensitivity $s^G$ for the constraint function $G$ was described already in Sec.~\ref{sec:shape-matching}.  The sensitivity for $M$ is $s^M_\alpha = \kappa_\alpha m(\bm{X}_\alpha) + n_\alpha^y \frac{\partial m(\bm{X}_\alpha)}{\partial Y_\alpha}$, where $\kappa_\alpha$ is the signed curvature and $n_\alpha^y$ is the $y$ component of the normal vector $\bm{n}_\alpha$.

Fig.~\ref{fig:dumb} shows results for both deterministic and stochastic optimisation for this problem.  The determinstic algorithm reveals that this is indeed a non-convex optimisation problem: that is, there are two local minima of the objective function.  Starting with a circular shape located in the upper lobe of the dumbbell, the deterministic algorithm converges quickly to a local optimum in which the shape is located close to the neck (2nd image in top row of Fig.~\ref{fig:dumb}).  The global optimum has the shape $\Omega$ inside the lower lobe, but convergence to this shape is frustrated because the near-circular locally-optimal shape cannot fit through the neck of the dumbbell.  In order to pass through the neck, while still obeying the constraint, the area inside the dumbbell must expand, to compensate the extra area outside.  However, this leads to an increase in the objective function $F$, which is not possible within the deterministic algorithm. 

However, on adding a weak stochastic element, one sees (for these parameters) that the system escapes the local optimum and converges to a steady state where it samples shapes that are close to the global optimum.  A final deterministic optimisation could be used to locate the true optimum, if required.  This example shows the potential usefulness of the stochastic level-set optimisation method, although this is obviously a very simple model at this stage.

\subsection{Compliance minimisation}
\label{sec:compliance}

As a final example, we show how this method can be applied to problems inspired by engineering applications of shape optimisation (this is distinct from topology optimisiation in that no holes are created during optimisation). As a proof of principle, we study a well-known benchmark problem of a two-dimensional cantilevered beam~\cite{Michell}, as shown in Fig.~\ref{fig:compliance}. 

In this problem, the grid used within the level-set method plays a second role as a finite-element mesh for the structure.  The objective function $F$ is the strain energy of the structure.  To calculate this, the structure is clamped at the left boundary ($L_x=0$) and a unit load is applied in the $y$-direction at position $(L_x,L_y/2)$.  The Young's modulus of the material inside $\Omega$ is $Y=100$, the Poisson's ratio is $0.3$.  The space outside $\Omega$ is occupied by a weak material whose density is $10^{-3}$.  The strain energy is minimised subject to the constraint that the total area of $\Omega$ is at least $0.5$ of the domain $\Omega_d$.  The sensitivity for the compliance is calculated by the method introduced in ~\cite{Dunning2015} and the sensitivity for the area constraint is \rlj{$s^G=-1$}, as discussed in Sec.~\ref{sec:shape-matching}.  The mesh size is $L_x\times L_y$ with $(L_x,L_y)=(40,20)$.  This relatively coarse mesh is used for computational convenience -- future work will exploit more efficient sensitivity calculations which will allow access to finer grids, but this is beyond the scope of this article.

Once the sensitivities are known, the method proceeds exactly as in the simple examples considered in previous sections.  The results in Fig.~\ref{fig:compliance} are qualitatively consistent with Fig.~\ref{fig:perimeter2}, and show that the stochastic level set method can be applied in such contexts.  The initial condition for the optimisation is a completely full grid $\Omega=\Omega_d$.  One sees from Figs.~\ref{fig:compliance}(a,b) that the optimiser first reduces the area of $\Omega$ in order to satisfy the constraint: during this part of the algorithm the strain energy increases.  Then, once the structure satisfies the constraint, its shape is optimised to reduce the strain energy.  Fig.~\ref{fig:compliance}(c) shows the optimal design that was found by deterministic optimisation, and a design found by the stochastic method.  Fig.~\ref{fig:compliance}(d) shows that increasing the noise increases the typical values of the objective function $F$, and the variance of this quantity also increases. However, Fig.~\ref{fig:compliance}(e) shows that the results are not quantitatively consistent with (\ref{equ:rwt-F}).  There are several potential sources of numerical error in this algorithm, of which the largest is expected to be the numerical uncertainties in calculations of sensitivities, due to the finite-element approximation.   We attribute the deviations from the predictions of (\ref{equ:rwt-F}) to these numerical uncertainties.

\begin{figure*}[tb]
\centering
\includegraphics[width=0.9\linewidth]{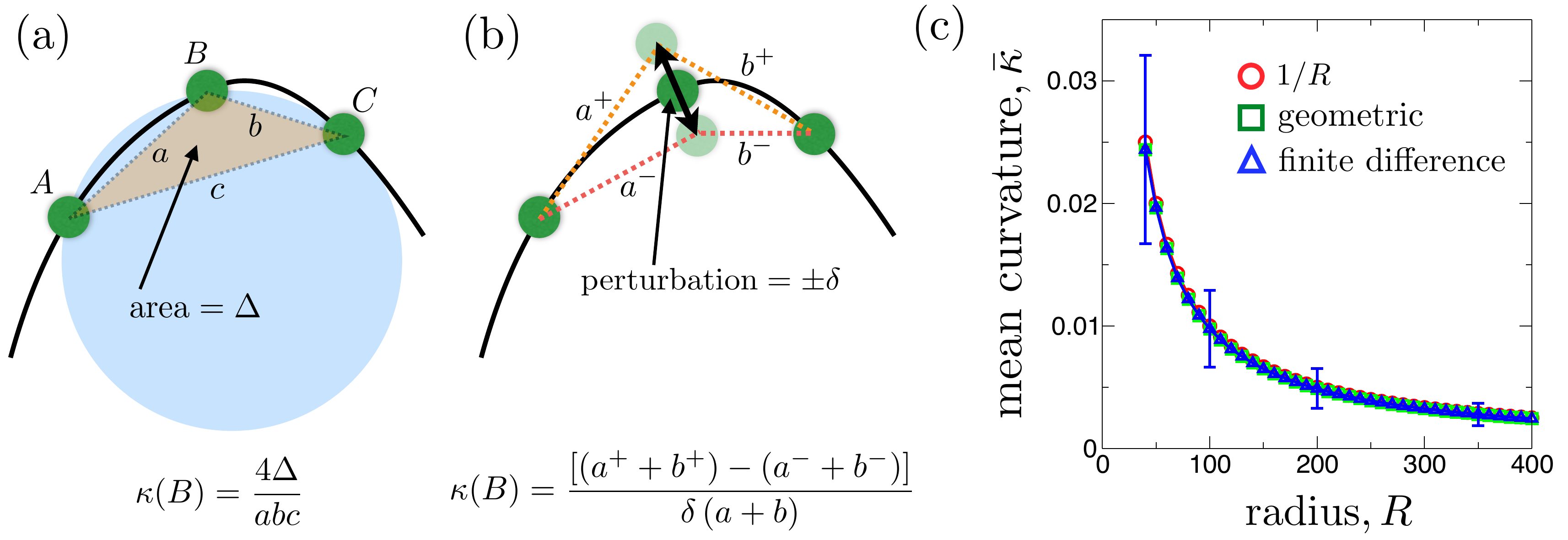}
\caption[Methods for estimating Euclidean curvature at a boundary point]
{\label{fig:curvature} Approximations for the local Euclidean curvature around a boundary point. (a) A simple geometric approximation can be obtained by matching a circle to any three sequential boundary points. The three points are assumed to lie on the perimeter of the circle, with the curvature defined as the inverse of the circle's radius \cite{Calabi1998}. (b) The curvature can also be calculated by performing an explicit finite-difference sensitivity calculation. A boundary point is displaced in the outward and inwards directions along its normal vector and a sensitivity is defined as the rate of change of the discretised perimeter per unit length, i.e. using a central finite-difference. In the limit of $\delta \rightarrow 0$ this defines a local perimeter sensitivity for the boundary point. (c)~Euclidean curvature for circles of increasing radius $R$. 
\rlj{The level-set is initialised as an exact signed-distance from a circle and the curvature is measured at each boundary point, using the geometric method from (a) and the finite-difference method from (b).  The mean curvature is obtained by averaging over the boundary points and is compared with the exact result $1/R$ (red circles).  For the finite-difference method we also show selected error bars whose size corresponds to the standard deviation of the measured curvatures, as discussed in the text.}
A value of $\delta = 10^{-4}$ was used for the central difference calculation.}
\end{figure*}

To understand this effect in more detail, it is useful to notice that (\ref{equ:rwt-F}) implies quite generally that $ \mathrm{Var}(F) = T^2\frac{\mathrm{d}}{\mathrm{d}T} \langle F \rangle$, where $\langle F \rangle$ is the mean strain energy at temperature $T$, and $\mathrm{Var}(F)=\langle F^2\rangle - \langle F \rangle^2$ is its variance.  This is an example of a fluctuation-dissipation theorem (FDT)~\cite{ChandlerBook}.  For the systems considered in Figs.~\ref{fig:bunny2},\ref{fig:perimeter2}, we find that this relation holds (otherwise the predictions based on the reweighting formula (\ref{equ:rwt-F}) would fail).  For this strain energy problem, we find that $\mathrm{Var}(F) \approx 3T^2 \frac{\mathrm{d}}{\mathrm{d}T} \langle F \rangle$, indicating that the variance of the objective is around three times larger than the prediction given by the FDT.  Our hypothesis is that these extra fluctuations in $F$ come from numerical errors associated with discretisation and with estimates of the sensitivity, but this remains an area for future study.

Finally we note that the globally-optimal structures for this (discretised) compliance problem are not simply-connected like the shapes considered in Fig.~\ref{fig:compliance}(c): the optimal structures include holes~\cite{Dunning2013}.  The stochastic method described here does not include an explicit prescription for the creation of holes during optimisation, even if these holes would reduce the objective function.  For this reason, we believe that the invariant measure of the method described here is of the form (\ref{equ:POmega}), with $p^0(\Omega)=0$ if $\Omega$ is not simply connected.  Generalisation of the method to include shapes with holes is an important area for future work.

\section{Conclusions}
\label{sec:conclusions}

We have introduced a stochastic level-set shape optimisation method, which is based on the deterministic (steepest-descent) method of~\cite{Dunning2015,Siva2016}.  The stochastic element of the algorithm acts on the boundary of the shape $\Omega$, and the method converges to a steady state in which it explores a range of shapes, according to a probability distribution (\ref{equ:POmega}).  The method is novel -- we are not able to prove rigorously that it converges to (\ref{equ:POmega}) but we have verified that this convergence does hold in two simple problems: shape matching (Fig.~\ref{fig:bunny2}) and perimeter minimisation at fixed area (Fig.~\ref{fig:perimeter2}).  A deeper mathematical analysis of the method would be in the scope of future work.

The motivation for introducing the stochastic method is to enable optimisation in non-convex problems, in which deterministic methods converge to local minima but are not able to search the whole parameter space in order to find the global optimum.  To demonstrate the idea, we have shown results in Fig.~\ref{fig:dumb} for a simple non-convex problem, in which the stochastic algorithm outperforms the deterministic one and converges to the global optimum.  For complex problems with multiple optima, an important feature of the method is that the underlying Boltzmann distribution in (\ref{equ:POmega}) allows it to be combined with methods such as parallel tempering, which are known to be effective in highly non-convex problems~\cite{frenkelsmit,Sugita1999,Rosta2009}.  Finally, we considered a model engineering problem (Fig.~\ref{fig:compliance}), for which we demonstrated that optimisation can be performed even for problems with complicated objective functions. The inherent numerical errors which are usually insignificant in deterministic optimisation are seen to be significant in the proposed stochastic optimisation and further work is needed to investigate this.

Compared with other stochastic optimisation methods that do not use gradient (or sensitivity) information~\cite{Wang2005,Wu2010,Luh2011}, the scheme presented here has two strengths.  First, as the noise strength is reduced, it recovers to the standard deterministic method, so it is guaranteed to peform no worse than that method (this property is not guaranteed in many stochastic algorithms).  Second, the fact that the invariant measure (or at least its $T$-dependence) is known to be (\ref{equ:POmega}) allows integration with parallel tempering and other methods from statistical physics~\cite{frenkelsmit,Sugita1999,Rosta2009}, which have proven extremely valuable in that context.

\section*{Acknowledgements}  
We thank Li-Tien Cheng for helpful discussions.  This article is based on work supported by the Air Force Office of Scientific Research, Air Force Material Command, USAF under Award No. FA9550-15-1-0316.

\begin{appendix}

\section{Sensitivity for the curve perimeter, local curvature and level-set reinitialisation}

\begin{figure*}[tb]
\centering
\includegraphics[width=0.8\linewidth]{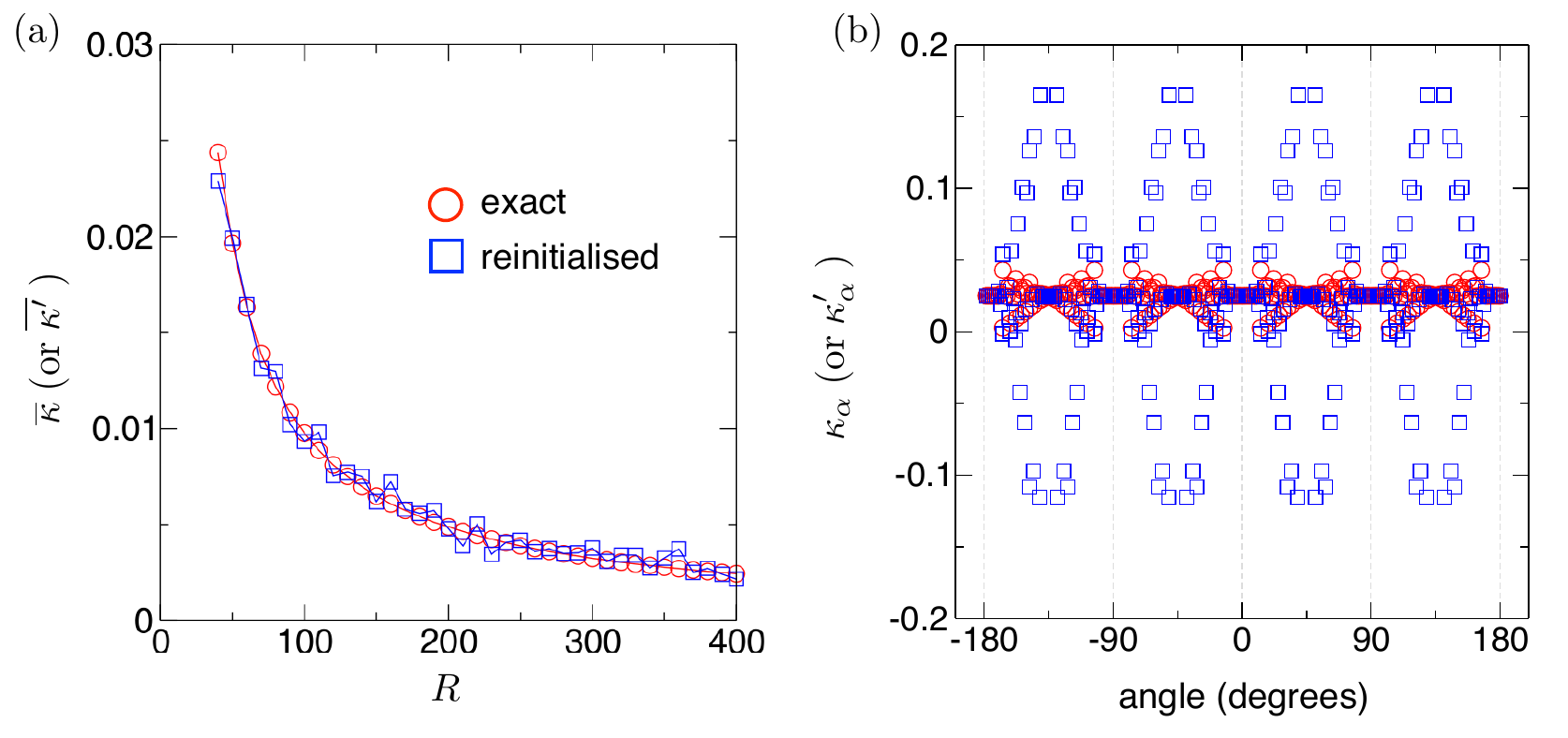}
\caption[Effects of discretisation on estimates of curvature]
{\label{fig:curvature2} Discretisation errors affect mean and local boundary point curvature. (a) Signed distance reinitialisation using a second-order Fast-Marching Method introduces errors into the mean boundary point curvature (blue squares). When the nodes of the level set domain are initialised using an exact signed-distance function (using the exact Euclidean distance to the interface) the mean curvature agrees near perfectly with the analytical result (red circles). In this case the only errors are introduced by the piece-wise linear discretisation of the boundary. (b) Local boundary point curvature around a circle of radius $R=40$. The angle is measured relative to the top of the circle. While the mean curvature is excellent, noise is present in the local boundary curvature, even when using a perfect signed-distance function (red circles). Reinitialisation of the signed-distance function leads to a significant increase in the noise (blue squares).}
\end{figure*}

\subsection{Local Curvature}

Several applications within this work make use of the sensitivity function associated with the perimeter of a curve.  The setting is illustrated in Fig.~\ref{fig:curvature}.  Recall the sensitivity is defined by (\ref{equ:sensF}). For a curve that is represented by a set of discrete boundary points, $s^F_\alpha$ can be estimated as shown in Fig.~\ref{fig:curvature}: one displaces a boundary point $\alpha$ by a distance $\delta$ in the direction normal to the curve, and calculates the change in $F$, which in this case is the perimeter.  It is convenient also to displace the boundary point by $-\delta$, which yields a central difference estimate of the sensitivity: for a boundary point with index $B$, this estimate is 
\begin{multline}
s^F_B = \frac{1}{2\ell_B\delta} \big[ F(\bm{X}_1,\ldots,\bm{X}_B+\bm{n}_\alpha\delta,\ldots,\bm{X}_n) 
 \\ 
- F(\bm{X}_1,\ldots,\bm{X}_B-\bm{n}_\alpha\delta,\ldots\bm{X}_n) \big]
\end{multline}
where $\ell_B$ is the length of the boundary segment associated with point $B$, and $F(\bm{X}_1,\ldots,\bm{X}_n)$ is the value of the objective function given the boundary point positions $(\bm{X}_1,\ldots,\bm{X}_n)$.

When the function $F$ is the perimeter of the curve, the sensitivity can be calculated as shown in Fig.~\ref{fig:curvature}b.  Using basic 2d geometry, one sees that this sensitivity can alternatively be obtained using the formula shown in Fig.~\ref{fig:curvature}a, and hence that this sensitivity is equal to the curvature $\kappa_B=\pm1/R$, where $R$ is the radius of the circle shown in Fig.~\ref{fig:curvature}(a): see also Ref.~\cite{Calabi1998}. 
The sign of $\kappa_B$ depends on the direction of the (inward) normal $\bm{n}$ at the point $B$: for a (locally) convex shape then $\kappa_B<0$ while for concave shapes $\kappa_B>0$.

It follows that for deterministic minimisation of shape perimeter in the absence of any constraint, one should recover curvature-driven flow of the boundary $\Gamma$ (effects of stochastic noise on this process have been considered by Souganidis and Yip~\cite{Souganidis2004}).  In our scheme, the computational implementation of this process is affected by the discretisation of the level-set field, and use of discrete boundary points.   In particular, this discretisation affects our estimation of sensitivities~\cite{Calabi1998}.  In Fig.~\ref{fig:curvature}(c), we initialise the level-set field as the signed distance from a circle of radius $R$ \rlj{with $\phi>0$ outside the circle. That is, the shape $\Omega$ includes the whole of $\Omega_{\rm d}$, except for a single circular hole, as in Fig.~\ref{fig:level-set}.}  We then infer the boundary points (as described in Sec.~\ref{sec:discrete-det}) and we measure the curvature $\kappa_\alpha$ at each such point, using the two methods shown in Fig.~\ref{fig:curvature}(a,b).  \rlj{The normal vectors $\bm{n}_\alpha$ point radially outwards on the perimeter of the circular hole, so the shape $\Omega$ is concave and one expects $\kappa_\alpha=1/R>0$.  The results in Fig.~\ref{fig:curvature}(c) show the mean curvature $\overline{\kappa}=(1/n)\sum_{\alpha=1}^n \kappa_\alpha$.  Error bars (for selected values of $R$) indicate the standard deviation $\sigma_\kappa = \sqrt{(1/n)\sum_{\alpha=1}^n (\kappa_\alpha-\overline{\kappa})^2}$.  If the boundary points were all located exactly on a circle of radius $R$ then one would have $\kappa_\alpha=1/R$ for all $\alpha$.  In practice, deriving the boundary points by linear interpolation from the level-set field leads to boundary points that are systematically slightly inside the circle, with offsets of order $a_0^2/R$ that depend on their position on the circle (recall that $a_0$ is the mesh size for the spatial discretisation of the level set field). This leads to a small systematic error in the measured mean curvature $\overline{\kappa}$ [relative error of order $(a_0/R)^2$] and a standard deviation $\sigma_\kappa$ that is approximately $\kappa/3$, independent of $R$ (see Fig.~\ref{fig:curvature}c).  See also Fig.~\ref{fig:curvature2}, which is discussed below.}


\subsection{Level-set reinitialisation and its effects on curvature measurements}

\rlj{In Fig.~\ref{fig:curvature2} we investigate curvature measurements in more detail.  In particular, we consider another discretisation effect, which arises from level-set reinitialisation.  Recall from Sec.~\ref{sec:levelset-update} that the level-set variables are periodically reinitialised, to maintain the property that $\phi_i$ is the signed distance of node $i$ from the boundary $\Gamma$.  
The reinitialisation procedure calculates the level-set variables $\phi_i$ as a function of the set of boundary point positions $\{\bm{X}_\alpha\}$.  However, if one starts with a set of boundary points, reinitialises the level set variables, and then recalculates a new set of boundary point positions $\{\bm{X}_\alpha'\}$, one does not have in general $\bm{X}_\alpha=\bm{X}_\alpha'$: that is, the boundary points are changed by the reinitialisation.}

This motivates the following numerical experiment, whose results are shown in Fig.~\ref{fig:curvature2}(a).  Start with a level set that encodes perfectly the signed distance from a circle of radius $R$.  Calculate the boundary points $\bm{X}_\alpha$ and their associated local curvatures $\kappa_\alpha$.  Then reinitialise the level set based on the boundary points $\bm{X}_\alpha$.  Based on this new level set, calculate a new set of boundary points $\bm{X}_\alpha'$ and their associated curvatures $\kappa_\alpha'$.  In the absence of discretisation errors, one would have $\kappa_\alpha = \kappa'_\alpha = 1/R$ for all $\alpha$.  
\rlj{Fig.~\ref{fig:curvature2}(a) shows averages of the $\kappa_\alpha$ and the $\kappa'_\alpha$, evaluated by summing over boundary points and dividing by the number of boundary points.  One has from Fig.~\ref{fig:curvature}c that the average $\overline{\kappa}$ is indeed close to $1/R$, but the average $\overline{\kappa'}$ shows significant numerical errors (deviations from $1/R$), which may be either positive or negative.

It was noted in the previous section that the estimated curvature varies with the position of the boundary point around the circle.  Fig.~\ref{fig:curvature2}(b) shows a representative set of curvature measurements ($\kappa_\alpha$ and $\kappa'_\alpha$), for a circle of radius $R=40$ (measured in units of $a_0$).  When the level set field is exactly equal to the signed distance from the circle, one sees variation in the curvatures $\kappa_\alpha$ that are of the same order as $\kappa$ itself, consistent with the error bars in Fig.~\ref{fig:curvature}c.  After reinitialisation of the level set field, one sees much larger errors.  The range of the $\kappa'_\alpha$ is large enough in this case that some values are even negative.  The source of this error is that reinitialisation leads to movements in the boundary points of order $1-5\%$ of a grid spacing.  The typical size of these movements depends weakly on the circle radius.  This leads to a standard deviation $\sigma_{\kappa'}$ that is approximately $0.06$, which is larger than the curvature itself, in contrast to $\sigma_{\kappa}\approx 0.008\approx\kappa/3$ when the level set is equal to the exact signed distance function.}


\subsection{Interpretation}

There are two conclusions from this analysis.  First, even if the level-set function is an accurate description of a shape (as with the analytic signed distance function considered here), one expects uncertainties in sensitivities, due to discretisation errors.  This can influence the convergence of the deterministic method to a minimum of $F$, and the extent to which the stochastic method samples (\ref{equ:POmega}).   Accurate estimation of sensitivities is therefore an important part of any future application of this method.  Second, reinitialisation of the level set can lead to small but significant movements of boundary points, which are large enough that local sensitivities change considerably.  In determinstic optimisation, the frequency of reinitialisation reduces as the system converges to the optimum, and this effect is not too pronounced.  On the other hand, in stochastic optimisation, reinitialisation is more frequent, and can affect the shapes $\Omega$ generated by the method.  A more detailed analysis of these effects will be included in future work.

\end{appendix}


\bibliographystyle{abbrv}
\bibliography{bib}

\end{document}